\documentclass[12pt]{article}

\renewcommand{\theequation}{\thesection.\arabic{equation}}

\newcommand{\shortoverline}{{\raisebox{1.4ex}{\rule{0.25cm}{0.01cm}}}{\hspace{-0.3cm}}}

\usepackage{amssymb}
\usepackage{graphicx}

\makeatletter
\def\eqnarray{%
 \stepcounter{equation}%
 \let\@currentlabel=\theequation
 \global\@eqnswtrue
 \global\@eqcnt\z@
 \tabskip\@centering
 \let\\=\@eqncr
 $$\halign to \displaywidth\bgroup\@eqnsel\hskip\@centering
 $\displaystyle\tabskip\z@{##}$&\global\@eqcnt\@ne
 \hfil$\displaystyle{{}##{}}$\hfil
 &\global\@eqcnt\tw@$\displaystyle\tabskip\z@{##}$\hfil
 \tabskip\@centering&\llap{##}\tabskip\z@\cr}
\makeatother

\begin{document}

\title{PSU(2,2$|$4) Exchange Algebra \\ of ${\cal N}$=4 Superconformal Multiplets  }

\author{Shogo Aoyama\thanks{e-mail: spsaoya@ipc.shizuoka.ac.jp}  \\
       Department of Physics \\
              Shizuoka University \\
                Ohya 836, Shizuoka  \\
                 Japan}

\maketitle

\begin{abstract}
 It is known that 
the unitary representation of the $D\hspace{-0.1cm}=\hspace{-0.1cm}4, \ {\cal N}\hspace{-0.1cm}=\hspace{-0.1cm}4$ superconformal multiplets and their descendants are constructed as  supercoherent states of bosonic and  fermionic  creation oscillators which   covariantly transform under SU(2,2$|$4). We non-linearly realize those  creation oscillators  on  the coset superspace PSU(2,2$|$4)/\{SO(1,4)$\otimes$SO(5)\} which is reparametrized by the $D\hspace{-0.1cm}=\hspace{-0.1cm}10$ supercoordinates $(X,\Theta)$. 
We consider  a $D=2$ non-linear $\sigma$ model on the coset superspace and set up  Poisson brackets for $X$ and $\Theta$ on the light-like line. It is then shown  that  the  non-linearly realized creation oscillators satisfy the classical exchange algebra with the classical r-matrix of PSU(2,2$|$4). 
We have recourse to purely algebraic quantization of the classical exchange algebra in which the r-matrix is promoted  to the universal R-matrix. The quantum exchange algebra  essentially characterizes correlation functions  of the $D\hspace{-0.1cm}=\hspace{-0.1cm}4,\ {\cal N}\hspace{-0.1cm}=\hspace{-0.1cm}4$ superconformal multiplets and their descendants  on the light-like line. It is because  they are supercoherent states of the oscillators. The arguments are straightforwardly extended to the case where those quantities  are endowed with the U($N$) YM gauge symmetry.

\end{abstract}

\vspace{2cm}


\newpage

\section{Introduction}
\setcounter{equation}{0}

The gauge/string duality between the $D=4, {\cal N}=4$ SUSY YM theory and the  IIB string theory on $AdS_5\times S^5$\cite{Ma}  is one of the subjects which have been discussed with great interest in recent years.  The integrability and the superconformal symmetry  PSU(2,2$|$4) play  crucial roles on  both the sides of the duality. 

The $D=4, \ {\cal N}=4$ SUSY YM theory on one side was casted to a $D=2$ spin-chain system with the superconformal symmetry PSU(2,2$|$4)\cite{Bei}. For this system the Bethe ansatz and the R-matrix  were extensively studied by assuming the integrability\cite{BeiStau}. However the origin of the integrability is obscure in this approach.  Moreover the existence of  the superconformal symmetry PSU(2,2$|$4) is also hypothetical, since it is broken by the Bethe ansatz to  two copies of the subgroup PSU(1,1$|$2) with  central charges. The appearance of  central charges makes the purely algebraic  construction of the universal  R-matrix for a simple group\cite{Pres,Kho} unreliable.  That is, the plug-in formula for the universal  R-matrix works only  if we  concern  a simple (super)group G and its Yangian generalization  Y(G)\cite{Pres,Kho}. This unusual feature of the R-matrix attracted  a particular interest as a challenging subject\cite{Bei2}. 
It gives a clue to study the anomalous scaling  dimension of  the ${\cal N}=4$ SUSY YM theory by means of the R-matrix of a $D=2$ spin-chain.

The IIB string theory on the other side was effectively described  by a $D=2$ non-linear $\sigma$-model on the coset superspace PSU(2,2$|$4)/\{SO(1,4)$\otimes$SO(5)\}\cite{Tsey}. It is integrable at the classical level admitting  an infinite  number of conserved currents\cite{Bena}.  Quantum extension of the integrability was argued by  the Bethe ansatz\cite{Aru}.  The Bethe ansatz is a common language to understand the gauge/string duality on    
 both of the sides. 
In this approach the origin of the integrability and the superconformal symmetry PSU(2,2$|$4) are clear because of the Poisson structure of the non-linear $\sigma$-model and the resemblance to the Green-Schwarz superstring respectively\cite{Ber}. 

In this paper we pursue the approach of the string side. For the non-linear $\sigma$-model on PSU(2,2$|$4)/\{SO(1,4)$\otimes$SO(5)\} we set up  Poisson brackets  for the basic fields  on the light-like line $x^+=const$ instead of the equal-time line $x^0=const$. In \cite{Ao1}  the consistency and the virtue for doing this  were shown for the non-linear $\sigma$-model on the general bosonic coset space
G/H. Namely, the Poisson brackets satisfy the three conditions. (i) They satisfies the Jacobi identities. 
(ii) The energy-momentum tensor $T_{--}$generates diffeomorphism on the light-like line by means of the Poisson brackets. 
(iii) At the origin of the coset space they coincide with  the Poisson brackets of the free boson theory. 
There exists a quantity $\Upsilon$, called  Killing scalar, which transforms as a linear representation vector of G by the Killing vectors of G/H. It exists in any representation of G and obeys the classical exchange algebra
\begin{eqnarray}
\{\Upsilon(x)\mathop{,}^\otimes \Upsilon(y)\}=-hr_{xy}\Upsilon(x)\otimes\Upsilon(y),    \label{CEA0}
\end{eqnarray}
on the light-like line with the Poisson brackets for the basic fields. Here $r_{xy}$ is the classical r-matrix of G. If G is a simple group, we may have recourse to the plug-in formula to promote it to the universal R-matrix $R_{xy}$, which is  expressed purely in terms of the generators of G\cite{Pres,Kho}. Then (\ref{CEA0}) becomes the quantum exchange algebra
\begin{eqnarray}
R_{xy}\Upsilon(x)\otimes \Upsilon(y)=\Upsilon(y)\otimes \Upsilon(x), \label{QEA0}
\end{eqnarray}
Its classical correspondence to (\ref{CEA0}) can be seen by 
$$
R_{xy}=1 +hr_{xy}+O(h^2). 
$$
 Correlation functions of $\Upsilon$s arrayed on the light-like line,  may be obtained by using the quantum exchange algebra to braid $\Upsilon$s at  adjacent positions successively.  

In this paper we apply all of these arguments to  the non-linear $\sigma$-model on PSU(2,2$|$4)/\ \{SO(1,4)$\otimes$SO(5)\}\cite{Tsey}. Now the basic fields of the coset space are the $D=10$  supercoordinates $(X,\Theta)$. For this non-linear $\sigma$-model the exchange algebra (\ref{CEA0}) or  (\ref{QEA0}) appears with the r- or R-matrix of the superconformal group PSU(2,2$|$4). 
The Killing scalar $\Upsilon$  is the ${\cal N}=4$ superconformal multiplet. But  PSU(2,2$|$4) is non-compact. Hence the  unitary representation contains infinitely many  descendants. We call them as a  whole the superconformal multiplet ${\cal V}$. In the previous work \cite{Ao1} the exchange algebra (\ref{CEA0}) or (\ref{QEA0}) of the non-linear $\sigma$-model on G/H was discussed in an arbitrary  representation. But the dimension of the unitary representation was  finite by assuming that G is a compact group. It is awkward to simply apply the arguments in \cite{Ao1} to the case where the dimension of  the unitary representation is necessarily infinite. It is our main concern to make a bridge over this gap.  

To this end we remember that the ${\cal N}=4$ superconformal multiplet ${\cal V}$ can  constructed over a supercoherent space of  bosonic and fermionic creation oscillators\cite{Bars}. 
The oscillators form a 8-d vector, say $\psi$, transforming covariantly under the superconformal group PSU(2,2$|$4). Let us write the covariant action on $\psi$ as an 8$\times$8 supermatrix  $e^{iM}$.  Then we show that the unitary representation is given by $\hat U=e^{i\bar\psi M\psi}$, which  is an infinite dimensional representation of PSU(2,2$|$4). 
 Acting  on  supercoherent states it induces the group action  $e^{iM}$ on the  8-d vector $\psi$, as shown by the state-operator relations (\ref{state-operator1}) and (\ref{state-operator2}).

Therefore the Killing scalar $\Upsilon$ which we  want to let satisfy the classical exchange algebra (\ref{CEA0})  is not necessarily the ${\cal N}=4$ superconformal multiplet $\cal V$, but may be the 8-d   covariant vector $\psi$. The Killing scalar $\Upsilon$ transforming identically with  $\psi$  can be readily constructed  on the coset space PSU(2,2$|$4)/\{SO(1,4)$\otimes$SO(5)\}, 
following \cite{Ao1}. Once this is done, the whole arguments in \cite{Ao1} can be applied to the non-linear $\sigma$-model on this coset space as well. That is, this Killing scalar $\Upsilon$ satisfies  the classical exchange algebra (\ref{CEA0})  with the r-matrix  in  the 8$\times$8 matrix representation of PSU(2,2$|$4). 
 PSU(2,2$|$4) is a simple group.\footnote {The R-/S-matrix discussed in \cite{Bei} is not the one for PSU(2,2$|$4), but for a non-simple group such as PSU(1,1$|$2)$\ltimes\mathbb{R}^3$. Further comments on this will be made at the end of this paper.} Hence the finite-dimensional r-matrix can be  quantized  to the universal R-matrix  by means of the  plug-in formula\cite{Pres,Kho}.  
Thus we get  the quantum  exchange algebra (\ref{QEA0}) for the Killing scalar $\Upsilon$ or equivalently 
for the covariant vector $\psi$. 
 From this we can calculate the  quantum  exchange algebra for  the ${\cal N}=4$ superconformal multiplet $\cal V$, because  $\cal V$ consists of $\psi$ as shown in table 2. The R-matrix for $\cal V$ is infinite-dimensional and yet algebraically the same as for the covariant vector $\psi$ owing to the operators-state relations (\ref{state-operator1}) and (\ref{state-operator2}). 
Thus we dispense with meeting the R-matrix in an infinite-dimensional representation head-on.

The paper is organized as follows. In section 2 we explain the ${\cal N}=4$ superconformal algebra of 
 PSU(2,2$|$4)  in terms of  bosonic and fermionic oscillators  forming  an 8-d covariant vector $\psi$. It is done by  following \cite{Bei} closely. In section 3 we construct the unitary representation of the superconformal group PSU(2,2$|$4) over a supercoherent space of the oscillators, following \cite{Bars}.  In particular we focus on the ${\cal N}=4$ field strength multiplet  appearing as a half-BPS state in  the unitary representation of  PSU(2,2$|$4),  which was discussed in \cite{Bei3}. Arguments on more general superconformal multiplets ${\cal V}$ are given in appendix A.  
The reader who is familiar the subjects may skip sections 2 and 3. In section  4 we discuss the 8$\times$8 supermatrix representation of PSU(2,2$|$4). The operator-state relations (\ref{state-operator1}) and  (\ref{state-operator2})
 establish a one-to-one map between the unitary(oscillator)  representation  in section 3 and 
 the matrix representation.  In section 5 the superconformal group PSU(2,2$|$4) is non-linearly realized on  the coset space PSU(2,2$|$4)/\{SO(1,4)$\otimes$SO(5)\}, in a way independent of the representation. Embedding the subgroup SO(1,4)$\otimes$SO(5) in PSU(2,2$|$4) is carefully studied. The salient feature of this coset space  is that 
 the basic fields of the coset space are the $D=10$ supercoordinates $(X,\Theta)$. In section 6 the oscillators, forming the 8-d covariant vector $\psi$ of PSU(2,2$|$4),  are non-linearly realized  on  PSU(2,2$|$4)/\{SO(1,4)$\otimes$SO(5)\} as the Killing scalar $\Upsilon$. In section 7,   we consider  the non-linear $\sigma$-model on the coset space and impose  Poisson brackets for $(X,\Theta)$,  according to \cite{Ao1}. Then we get the classical exchange algebra for the non-linearly realized  oscillators $\Upsilon$ and discuss 
its implication for correlation functions  when the non-linear $\sigma$-model 
is quantized on the light-like line. Appendix A is devoted to complete the argument on the unitary(oscillator)  representation of  PSU(2,2$|$4) in section 3.
 Superconformal multiplets other than the field strength multiplet appear as larger BPS multiplets. 
 Though they were argued in various works \cite{Bei3,Doh, Ferr}, here we straighten  the arguments  by unifying the notations. Finally in appendix B we explain how to calculate the Killing vectors of the general coset space G/H in a way independent of the representation, i.e., by using only the Lie-algebra.  The unitary(oscillator) representation of PSU(2,2$|$4) as well as  the  matrix one  require  central charges as shown in section 3 and 4. The algebraic calculation in appendix B dispenses us with meeting central charges. It is desirable since  PSU(2,2$|$4) is a simple group which is free from central charges at the algebraic level and so are the Killing vectors.

\section{The ${\cal N}$=4 SUSY YM theory and  PSU(2,2$|$4)}
\setcounter{equation}{0}

The ${\cal N}$=4 SUSY YM theory is described by a set of fundamental fields 
\begin{eqnarray}
      A_\mu, \Psi_{\alpha a}, \Psi_{\dot\alpha}^{\ a}, \Phi_{[a,b]}.
 \nonumber
\end{eqnarray}
Our index convention is as follows: $\mu$ refers to vector indices of the Lorentz group SO(1,3), taking four values. $\alpha, \dot\alpha$ refer to two independent spinor indices of SU(2)$\otimes$SU(2)($\in$ SU(2,2)). They respectively takes two values. $a,b$ refer spinor indices of the R-symmetry SU(4), taking four values. $[\ ,\ ]$ indicates anti-symmetrization of them. Complex conjugation of the spinor representation is indicated by raising or lowering    
  indices. 
The ${\cal N}$=4 SUSY field strength  multiplet is constructed out of these fundamental fields as shown in table 1. There $F$ indicates the field strength $F_{\mu\nu}$, which has been  split into $F_{\{\alpha\beta\}}$ and $F_{\{\dot\alpha\dot\beta\}}$ by using the spinor indices of SU(2)$\otimes$SU(2). \{\ ,\ \} indicates symmetrization of the indices. ${\frak D}$ indicates space-time derivative ${\frak D}_\mu$, which may be written   as ${\frak D}_{\alpha\dot\beta}$.
 The representation of SU(2)$\otimes$SU(2) and SU(4) are indicated by the Dynkin labels for the highest weight as [$s_1,s_2$] and [$r_1$,$r_2$,$r_3$] respectively. 
  The Young tableau representing the SU(4) representation  is drawn in figure 1. Their dimensions are given by
\begin{eqnarray}
dim [s_1,s_2] &=& {\bar s}_1{\bar s}_2,  \nonumber \\
dim [r_1,r_2,r_3] &=& {1\over 12}{\bar r}_1{\bar r}_2{\bar r}_3
({\bar r}_1+{\bar r}_2)({\bar r}_2+{\bar r}_3)({\bar r}_1+{\bar r}_2+{\bar r}_3),  \nonumber 
\end{eqnarray}
with ${\bar s}_i=s_i+1$ and ${\bar r}_a=r_a+1$.

\vspace{0.5cm}

\newcommand{\lw}[1]{\smash{\lower1.5ex\hbox{#1}}}

\begin{table}[ht]
\begin{center}
\begin{tabular}{|c|cc|cc|}  \hline
\lw {field}  & \multicolumn{2}{c|}{SU(2)$\otimes$SU(2)}   &  \multicolumn{2}{c|}{SU(4)}       \\
\vspace{-0.6cm}                         \\
 & \multicolumn{2}{c|}{h.w.}   &  \multicolumn{2}{c|}{h.w.}       \\
\hline 
${\frak D}^k F$  & $\hspace{0.4cm}[k+2,k]$    && \hspace{0.2cm} [0,0,0] &  \\
${\frak D}^k \Psi$ & $\hspace{0.4cm}[k+1, k]$  && \hspace{0.2cm} [1,0,0] &  \\
${\frak D}^k \Phi$  &\hspace{0.4cm}  $[k,k]$ && \hspace{0.2cm} [0,1,0] & \\
${\frak D}^k \dot\Psi$ &\hspace{0.4cm} $[k, k+1]$ &&\hspace{0.2cm} [0,0,1] & \\
${\frak D}^k \dot F $ &\hspace{0.4cm} $[k,k+2]$ &&\hspace{0.2cm} [0,0,0]  & \\
\hline
\end{tabular}

\caption{${\cal N}=4$ SUSY field strength multiplet.}

\end{center}
\end{table}

\vspace{-0.5cm}

\begin{figure}[h]

\setlength{\unitlength}{1mm}
\begin{picture}(70,28)(-40,-20)

\put(70,7){\line(0,-1){7}}
\put(63,7){\line(0,-1){7}}
\put(56,7){\line(0,-1){7}}
\put(49,7){\line(0,-1){14}}
\put(42,7){\line(0,-1){14}}
\put(35,7){\line(0,-1){14}}
\put(28,7){\line(0,-1){21}}
\put(21,7){\line(0,-1){21}}
\put(14,7){\line(0,-1){21}}
\put(7,7){\line(0,-1){21}}
\put(0,7){\line(0,-1){21}}

\put(0,7){\line(1,0){70}}
\put(0,0){\line(1,0){70}}
\put(0,0){\line(1,0){49}}
\put(0,-7){\line(1,0){49}}
\put(0,-14){\line(1,0){28}}

\put(14,0){\makebox(7,7){$\cdots$}}
\put(35,0){\makebox(7,7){$\cdots$}}
\put(56,0){\makebox(7,7){$\cdots$}}
\put(14,-7){\makebox(7,7){$\cdots$}}
\put(35,-7){\makebox(7,7){$\cdots$}}
\put(14,-14){\makebox(7,7){$\cdots$}}

\put(0,-15.5){
$\underbrace{\makebox(25,3){}  }_{\mbox{$r_3$}}$
}

\put(28,-8.5){
$\underbrace{\makebox(18,3){}  }_{\mbox{$r_2$}}$
}

\put(49,-1.5){
$\underbrace{\makebox(18,3){}  }_{\mbox{$r_1$}}$
}
\end{picture}

\caption{The SU(4) Young tableau for the representation with the Dynkin label [$r_1$,$r_2$,$r_3$].}

\end{figure}
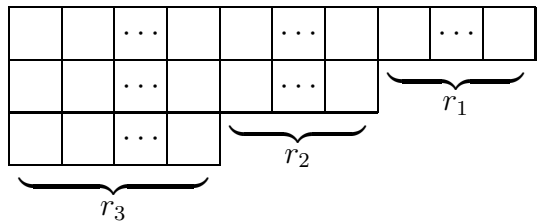

\vspace{0.5cm}

The ${\cal N}$=4 SUSY YM theory has  the  superconformal symmetry defined by the supergroup PSU(2,2$|$4). 
It is represented as a subgroup of  the  slightly enlarged supergroup U(2,2$|$4). The Lie-algebra of U(2,2$|$4)  is decomposed as
\begin{eqnarray}
T^-\oplus T^0 \oplus T^+,  \nonumber
\end{eqnarray}
where $T^0$ represents generators of the compact subgroup U(2,2)$\otimes$U(4)$\otimes$U(1) and $T^- \oplus T^+$ represents  non-compact ones such that 
$$
[T^0,T^\pm]=T^\pm,\quad\quad  [T^\pm,T^\pm\}=T^\pm.
$$
Here the bracket 
 $[\ ,\ \}$  is a graded commutator understood as an anti-commutator between fermionic generators, and as a commutator otherwise. 
 We introduce 
two set of bosonic oscillators $(a_\alpha,a^{\dagger\alpha})$ and $(b_{\dot\alpha},b^{\dagger\dot\alpha})$ and one set of fermionic ones $(c_a,c^{\dagger a})$
 to  realize these generators. The non-trivial  commutation relations  are 
\begin{eqnarray}
[a^\alpha,a^\dagger_\beta]=\delta^\alpha_\beta,\quad\quad
[b^{\dot\alpha},b^\dagger_{\dot\beta}]=\delta^{\dot\alpha}_{\dot\beta},\quad\quad 
\{c^a,c^\dagger_b\}=\delta^a_b.  \label{commutation}
\end{eqnarray}
To be explicit,   $T^0$ consists of the generators 
\begin{eqnarray}
L^\alpha_{\ \beta} &=& a^\dagger_\beta a^\alpha-{1\over 2}\delta_\beta^\alpha a^\dagger_\gamma a^\gamma,  \nonumber \\
L^{\dot\alpha}_{\ \dot\beta} &=& b^\dagger_{\dot\beta} b^{\dot\alpha}-{1\over 2}\delta_{\dot\beta}^{\dot\alpha} b^\dagger_{\dot\gamma} b^{\dot\gamma}, \label{LLR} \\
R^a_{\ b}&=&c^\dagger_b c^a-{1\over 4}\delta^a_b c^\dagger_c c^c, \nonumber
\end{eqnarray}
and the three U(1) generators   
\begin{eqnarray}
D&=&\ \ 1+{1\over 2}a^\dagger_\gamma a^\gamma+ {1\over 2}b^\dagger_{\dot\gamma} b^{\dot\gamma}={1\over 2}a^\dagger_\gamma a^\gamma+ {1\over 2} b^{\dot\gamma} b^\dagger_{\dot\gamma},  \nonumber\\
C&=&\ \ 1- {1\over 2}a^\dagger_\gamma a^\gamma+ {1\over 2}b^\dagger_{\dot\gamma} b^{\dot\gamma}-{1\over 2}c^\dagger_c c^c=
- {1\over 2}a^\dagger_\gamma a^\gamma+ {1\over 2}b^{\dot\gamma} b^\dagger_{\dot\gamma} -{1\over 2}c^\dagger_c c^c,  \label{U(1)}\\
B&=&-1+ {1\over 2}a^\dagger_\gamma a^\gamma- {1\over 2}b^\dagger_{\dot\gamma} b^{\dot\gamma}=
 {1\over 2}a^\dagger_\gamma a^\gamma- {1\over 2}b^{\dot\gamma} b^\dagger_{\dot\gamma}.  \nonumber
\end{eqnarray}
The generators in $T^+$ are given by 
\begin{eqnarray}
Q^a_{\ \alpha}&=& a^\dagger_\alpha c^a, \quad\quad
{\dot Q}_{\dot\alpha a}=b^\dagger_{\dot\alpha }c^\dagger_a, \quad\quad 
P_{\dot\alpha \beta}= b^\dagger_{\dot\alpha} a^\dagger_{\beta}, \nonumber
\end{eqnarray}
while those in $T^-$  by \footnote{If $Q^a_{\ \alpha}= a^\dagger_\alpha c^a, S^\alpha_{\ a} = c^\dagger_a a^\alpha$ are replaced by $Q_{a \alpha}= a^\dagger_\alpha c^\dagger_a, S^{\alpha a} = c^a a^\alpha$ , they form the algebra 
$$
[T^0,T^\pm]=T^\pm,\quad\quad \{T^+,T^-\}=T^0,\quad\quad  \{T^\pm,T^\pm\}=0.
$$
This form of the Lie-algebra U(2,2$|$4)  was used to discuss the unitary representation in refs \cite{Gnu} } 
\begin{eqnarray}
S^\alpha_{\ a} &=& c^\dagger_a a^\alpha \quad\quad 
{\dot S}^{\dot\alpha a}= b^{\dot\alpha} c^a, \quad\quad
K^{\alpha\dot\beta}=a^\alpha b^{\dot\beta}. \nonumber
\end{eqnarray}
Then the generators in (\ref{LLR}) form the subalgebra  SU(2)$\otimes$SU(2)$\otimes$SU(4) of U(2,2$|$4) 
\begin{eqnarray}
\ [L^\alpha_{\ \beta}, L^\gamma_{\ \delta}]&=& -\delta_\beta^\gamma L^\alpha_{\ \delta}+\delta^\alpha_\delta L^\gamma_{\ \beta}, \quad\quad 
[L^{\dot\alpha}_{\ \dot\beta}, L^{\dot\gamma}_{\ \dot\delta}]= -\delta_{\dot\beta}^{\dot\gamma} L^{\dot\alpha}_{\ \dot\delta}+\delta^{\dot\alpha}_{\dot\delta} L^\gamma_{\ \beta},  
  \nonumber\\
&\ & \hspace{1cm}[R^a_{\ b}, R^c_{\ d}]=-\delta_b^c R^a_{\ d}+\delta^a_d R^c_{\ d}.    \label{L0}
\end{eqnarray}
The algebra $[T^\pm,T^\pm\}=T^\pm$ is nilpotent in the sense that $[T^\pm[T^\pm,T^\pm\}\}=0$,  and is given by 
\begin{eqnarray}
\{{\dot Q}_{\dot\alpha b}, Q^a_{\ \beta}\}=\delta^a_b P_{\dot\alpha\beta},
 \quad\quad
\{{\dot S}^{a\dot\beta},S^\alpha_{\ b}\}=\delta^a_bK^{\beta\dot\alpha},
\label{L++}
\end{eqnarray} 
while the algebra $[T^+,T^-]=T^0$  by
\begin{eqnarray}
\ [K^{\alpha\dot\beta}, P_{\dot\gamma\delta}] &=&\delta^{\dot\beta}_{\dot\gamma}L^\alpha_{\ \delta} +\delta^\alpha_{\delta}{\dot L}^{\dot\beta}_{\ \dot\gamma}    + \delta^{\dot\beta}_{\dot\gamma}\delta^\alpha_{\delta}D,  \nonumber\\
\{S^\alpha_{\ b},Q^a_{\ \beta}\}&=& \delta^a_b L^\alpha_{\ \beta}+\delta^\alpha_{\beta}R^a_{\ b} + {1\over 2}\delta^a_b\delta^\alpha_{\beta}(D-C),    \label{L+-}\\
\{{\dot S}^{a\dot\beta}, {\dot Q}_{\dot\alpha b}\}&=& \delta^a_b{\dot L}^{\dot\beta}_{\ \dot\alpha} - \delta_{\dot\alpha}^{\dot\beta}R^a_{\ b}+{1\over 2}\delta^a_b\delta_{\dot\alpha}^{\dot\beta}(D+C).   \nonumber 
\end{eqnarray}
Finally the algebra $[T^+,T^-]$, which does not close into $T^0$,  is given by 
\begin{eqnarray}
[S^\alpha_{\ b},P_{\dot\alpha\beta}]&=&\delta^\alpha_\beta{\dot Q}_{\dot\alpha b},
\ \quad\quad [K^{\alpha\dot\beta},Q_{\dot\alpha b}]=\delta^{\dot\beta}_{\dot\alpha}S^\alpha_{\ b},   \nonumber \\ 
\ [{\dot S}^{a\dot\beta},P_{\dot\alpha\beta}]&=& \delta_{\dot\alpha}^{\dot\beta}Q^a_{\ \beta}, \quad\quad [K^{\alpha\dot\beta}, Q^a_{\ \beta}]=\delta^\alpha_\beta {\dot S}^{a\dot\beta}.    \label{L0+} 
\end{eqnarray}
We omit the algebra  $[T^0,T^\pm]=T^\pm$, which can be easily written down. 
  Altogether the algebrae (\ref{L0})$\sim$(\ref{L0+}) define the Lie-algebra of U(2,2$|$4)\cite{Bei}. 

It is instructive to put the generators in  a tensor product form of the row and column vectors
\begin{eqnarray}
\psi^\dagger = (a^\dagger_\beta,c^\dagger_b,b^{\dot\beta}), \quad\quad\quad
\psi=
\left(
\begin{array}{c}
a^\alpha \\
c^a  \\
b^\dagger_{\dot\alpha}
\end{array}
\right),    \label{psi}
\end{eqnarray}
as
\begin{eqnarray}
{\scriptsize
\psi^\dagger\otimes \psi=\left[
\begin{array}{c|c|c}
\hspace{0.2cm}L^\alpha_{\ \beta} \hspace{0.2cm} &\hspace{0.2cm}S^\alpha_{\ b}  \hspace{0.2cm}&\hspace{0.2cm} K^{\alpha\dot\beta} \hspace{0.2cm}  \\
\vspace{-0.2cm}     &    &       \\
\hline
\vspace{-0.2cm}  &    &    \\
\hspace{0.2cm}Q^{a}_{\ \beta} \hspace{0.2cm} & \hspace{0.2cm}R^{a}_{\ b} \hspace{0.2cm} & \hspace{0.2cm} {\dot S}^{a \dot\beta}  \hspace{0.2cm}  \\
\vspace{-0.2cm}     &      &     \\
\hline
\vspace{-0.2cm}  &   &    \\
\hspace{0.2cm}P_{\dot\alpha\beta} \hspace{0.2cm}  &\hspace{0.2cm}{\dot Q}_{\dot\alpha b}  \hspace{0.2cm} & \hspace{0.2cm} {\dot L}_{\ \dot\alpha}^{\dot\beta}  \hspace{0.2cm}  
\end{array}
\right]} +
{\scriptsize
\left[
\begin{array}{c|c|c}
{1\over 2}\delta^\alpha_{\beta}(D+B)  &0  & 0  \\
\vspace{-0.2cm}     &    &       \\
\hline
\vspace{-0.2cm} &    &    \\
0  & -{1\over 2}(C+B)  & 0   \\
\vspace{-0.2cm}     &      &     \\
\hline
\vspace{-0.2cm}  &   &    \\
0  &0   &  {1\over 2}\delta_{\dot\alpha}^{\dot\beta}(D-B)  
\end{array}
\right].}\ \ \ \  \label{matrix}
\end{eqnarray}
Here use was made of (\ref{U(1)}). $P_{\dot\alpha\beta}, Q_{\ \beta}^{ a}, {\dot Q}_{\dot\alpha b }$ in the lower-left blocks are generators of (super)\ translation
 while $K^{\alpha\dot\beta}, S_{\ b}^{\alpha}, {\dot S}^{a\dot\beta}$ in the upper-right blocks are generators of (super)\ boost. In the diagonal blocks 
$L^\alpha_{\ \beta}, {\dot L}^{\dot\beta}_{\ \dot\alpha}, R^a_{\ b}$ are 
generators of the Lorentz subsymmetry SU(2)$\otimes$SU(2)($\in$SU(2,2))  and the R-symmetry SU(4), and $D,C,B$ are three U(1) charges. $D$ is the dilatation. $B$ never appears in the above superalgebrae of U(2,2$|$4), (\ref{L0})$\sim$(\ref{L0+}).  All the generators commute with $C$. Hence $C$ is a central charge.

Finally  we get the quadratic Casimir in the form 
\begin{eqnarray}
L^\alpha_{\ \beta} L^\beta_{\ \alpha}-R^a_{\ b}R^b_{\ a}+L^{\dot\alpha}_{\ \dot\beta} L^{\dot\beta}_{\ \dot\alpha}+D^2 -\{P_{\dot\alpha\beta},K^{\beta\dot\alpha}\}-
[Q^a_{\ \alpha},S^\alpha_{\ b}]-[{\dot Q}_{\dot\alpha a},{\dot S}^{a\dot\alpha}],  \label{Casimir0}
\end{eqnarray}
as can be checked by a direct calculation.

\section{Unitary(oscillator) representation of  PSU(2,2$|$4)}
\setcounter{equation}{0}

The superconformal transformations act on the ${\cal N}=4$ SUSY field strength multiplet given in table 1. In quantum field theory they are represented as  unitary linear transformations in the Hilbert space.  Hence  the unitary representation of the superconformal group PSU(2,2$|$4) is the primary concern for quantization   of  the ${\cal N}$=4 SUSY YM theory. 
Since PSU(2,2$|$4) is non-compact , the unitary representation is necessarily infinite-dimensional. The ${\cal N}$=4 SUSY field strength multiplet is one of infinitely many multiplets in the unitary representation of PSU(2,2$|$4). Other  multiplets, generally called ${\cal N}$=4 superconformal  multiplets, are known by a systematic analysis of the unitary representation\cite{Bei3,Doh, Ferr}.  They are given in appendix A. 

A unitary operator $\hat U$ representing U(2,2$|$4)  may be  given by 
\begin{eqnarray}
\hat U=e^{i\bar\psi M \psi},    \label{uop}
\end{eqnarray}
with $\bar\psi=\psi^\dagger\gamma$\cite{Bars}. Here $M$ and $\gamma$ are $8\times 8$ supermatrices of the block form 
\begin{eqnarray}
M=\left(
\begin{array}{c|c|c}
V  &\hspace{0.2cm}\theta  \hspace{0.2cm}&\hspace{0.2cm} X \hspace{0.2cm}  \\
\vspace{-0.3cm}     &    &       \\
\hline
\vspace{-0.3cm}  &    &    \\
\hspace{0.2cm}\theta^\dagger  & \hspace{0.2cm}W \hspace{0.2cm} & \ \epsilon  \hspace{0.2cm}  \\
\vspace{-0.3cm}     &      &     \\
\hline
\vspace{-0.3cm}  &   &    \\
-X^\dagger   &\hspace{0.2cm}-\epsilon^\dagger  \hspace{0.2cm} & \hspace{0.2cm} Z  \hspace{0.2cm}  
\end{array}
\right),  
\quad\quad\quad\quad 
\gamma=\left(
\begin{array}{c|c|c}
1  &\hspace{0.2cm}0  \hspace{0.2cm}&\hspace{0.1cm} 0 \hspace{0.1cm}  \\
\vspace{-0.3cm}     &    &       \\
\hline
\vspace{-0.3cm}  &    &    \\
\hspace{0.2cm}0  & \hspace{0.2cm}1 \hspace{0.2cm} & \ 0  \hspace{0.1cm}  \\
\vspace{-0.3cm}     &      &     \\
\hline
\vspace{-0.3cm}  &   &    \\
0   &\hspace{0.2cm}0  \hspace{0.2cm} &  -1   
\end{array}
\right), 
 \label{M}
\end{eqnarray}
in which $V,W,Z$ are Hermitian matrices, $X$ is a complex matrix, but 
$\theta$(or $\epsilon$) is a 2$\otimes$4(or 4$\otimes$2) matrix of which elements are Grassmannian numbers. The unitarity of $\hat U$ follows from the Hermiticity of $\bar\psi M\psi$, i.e., $(\bar\psi M\psi)^\dagger =\bar\psi M\psi$. The vector $\psi$ transforms covariantly by the action of U(2,2$|$4) as 
\begin{eqnarray}
{\hat U}^\dagger \psi {\hat U} =e^{iM}\psi  \label{state-operator1},
\end{eqnarray}
and $\bar\psi$ contravariantly as 
\begin{eqnarray}
{\hat U}^\dagger \bar\psi {\hat U} =\bar\psi e^{-iM}  \label{state-operator2}.
\end{eqnarray}
The {\it minus} sign in $M$ is a hallmark of  non-compactness of U(2,2$|$4). It comes from the fact that we have chosen $\psi$ in (\ref{psi}) as  having creation and  annihilation oscillators mixed. 
For representing the compact supergroup U(4$|$4), it suffices to define $\psi$ by annihilation oscillators alone. Consequently 
the {\it minus} sign  is not needed for the block matrices $X^\dagger$ and $\epsilon^\dagger$ in  $M$. Then $M$ is Hermitian in itself and  $\gamma$ is not needed either. 

We explain this point of the unitary(oscillator) representation by taking much simpler groups SU(1,1) and SU(2) as  examples. Both Lie-algebrae are realized by using two pairs of oscillators $(a,a^\dagger)$, and $(b,b^\dagger)$. The non-trivial commutation relations  are
\begin{eqnarray}
[a,a^\dagger] =1,\quad\quad [b,b^\dagger] =1.  \nonumber
\end{eqnarray}
Then SU(1,1) is realized by  the unitary operator (\ref{uop}) with  
\begin{eqnarray}
\psi^\dagger = (a^\dagger,b), \quad\quad
\psi=
\left(
\begin{array}{c}
a \\
b^\dagger
\end{array}
\right),  
\quad
M=\left(
\begin{array}{cc}
 V& \phi \\
 -\phi^*&-V 
\end{array}\right), \quad\quad
\gamma=\left(
\begin{array}{cc}
 1& 0 \\
 0&-1 
\end{array}\right),   \nonumber
\end{eqnarray}
while SU(2) by  the unitary operator $e^{i{\psi_{\mbox{\tiny U}}}^\dagger M_{\mbox{\tiny U}}\psi_{\mbox{\tiny U}}}$ with 
\begin{eqnarray}
{\psi_{\mbox{\tiny U}}}^\dagger = (a^\dagger,b^\dagger), \quad\quad
\psi_{\mbox{\tiny U}}=
\left(
\begin{array}{c}
a \\
b
\end{array}
\right),  
\quad
M_{\mbox{\tiny U}}=\left(
\begin{array}{cc}
 V& \phi \\
 \phi^*& -V 
\end{array}\right). \nonumber \hspace{4.2cm}
\end{eqnarray}
From $\bar\psi M\psi$ and ${\psi_{\mbox{\tiny U}}}^\dagger M_{\mbox{\tiny U}}\psi_{\mbox{\tiny U}}$ we read the generators of the respective group as
\begin{eqnarray}
{\rm SU(1,1)}&:&\hspace{0.8cm} T^+=a^\dagger b^\dagger, \quad T^-=a b, \quad\ T^0={1\over 2}(a^\dagger a+ b^\dagger b),   \nonumber \\
{\rm SU(2)}&:&\hspace{0.7cm} {T_{\mbox{\tiny U}}}^+=a^\dagger b, \quad {T_{\mbox{\tiny U}}}^-=a b^\dagger, \quad {T_{\mbox{\tiny U}}}^0={1\over 2}(a^\dagger a- b^\dagger b), \nonumber 
\end{eqnarray}
which satisfy the algebrae
$$
\ [T^+,T^-]=-2T^0,\quad\quad [T^0,T^\pm]=\pm T^\pm,   
$$
and
$$
\ [T_{\mbox{\tiny U}}^+,T_{\mbox{\tiny U}}^-]=2T_{\mbox{\tiny U}}^0,\quad\quad [T_{\mbox{\tiny U}}^0,T_{\mbox{\tiny U}}^\pm]=\pm T_{\mbox{\tiny U}}^\pm. 
$$
Let $|0>$ to be the vacuum of the Fock space. Then we have 
$$
(T^+)^n a^\dagger|0>\ne 0,\quad\quad (T_{\mbox{\tiny U}}^+)^n a^\dagger|0>= 0,
$$
for a positive integer $n$. Thus by means of the unitary operator (\ref{uop}) we can realize the non-compact group U(1,1) in an infinite dimensional representation.

We return to the main arguments on  PSU(2,2$|$4). 
The unitary operator  (\ref{uop}) for PSU(2,2$|$4) acts on a Fock space given by all possible oscillator excitations
\begin{eqnarray}
\prod_{\alpha=1}^2\prod_{n_{a_\alpha}=1}^\infty(a^\dagger_\alpha)^{n_{a_\alpha}}
\prod_{\dot\alpha=1}^2\prod_{n_{b_{\dot\alpha}}=1}^\infty(b^\dagger_{\dot\alpha})^{n_{b_{\dot\alpha}}}
\prod_{a=1}^4\prod_{n_{c_{a}}=1}^\infty(c^\dagger_{a})^{n_{c_a}}|0>.
  \nonumber
\end{eqnarray}
Thus it is the unitary representation of U(2,2$|$4). 
PSU(2,2$|$4) is represented in a subsector of the Fock space constrained by
\begin{eqnarray}
C=1-{1\over 2}\sum_{\alpha}^2n_{a_\alpha} +{1\over 2}\sum_{\dot\alpha}^2n_{b_{\dot\alpha}} -{1\over 2}\sum_{a}^4n_{c_a}=0,  \label{C}
\end{eqnarray}
with $C$ given in  (\ref{U(1)}). If we have 
\begin{eqnarray}
a^\alpha|0>=0,\quad\quad b^{\dot\alpha}|0>=0\quad\quad 
c^a|0>=0, \nonumber
\end{eqnarray}
the vacuum $|0>$ is not in this subsector because $C=1$. Hence  
we  define a new physical vacuum $Z$ which has $C=0$. It may be realized by 
\begin{eqnarray}
 Z=c^\dagger_3c^\dagger_4|0>.        \label{ZZ}
\end{eqnarray}
It is convenient to rename  the whole fermionic oscillators $c^a,c^\dagger_a, a=1,2,3,4$ as \cite{Bei3}
\begin{eqnarray}
(c^1,c^2)&\equiv& c^{\bar a}, \quad\quad  (c^3,c^4)=(d^\dagger_3,d^\dagger_4)\equiv d^\dagger_{\dot a},   \nonumber \\
(c^\dagger_1,c^\dagger_2)&\equiv&c^\dagger_{\bar a},\quad\quad 
(c^\dagger_3,c^\dagger_4)=(d^3,d^4)\equiv d^{\dot a}.
\label{renamee}
\end{eqnarray}
Then $Z$ satisfies 
\begin{eqnarray}
a^\alpha|0>=0,\quad\quad b^{\dot\alpha}|0>=0,\quad\quad  
c^{\bar a}|0>=0, \quad\quad  d^{\dot a}|0>=0. \nonumber
\end{eqnarray}
The physical Fock space is built up on this $Z$ as 
\begin{eqnarray}
\prod_{\alpha=1}^2(a^\dagger_\alpha)^{n_{a_\alpha}}
\prod_{\dot\alpha=1}^2(b^\dagger_{\dot\alpha})^{n_{b_{\dot\alpha}}}
\prod_{\bar a=1}^2(c^\dagger_{\bar a})^{n_{c_{\bar a}}}\prod_{\dot a=3}^4(d^\dagger_{\dot a})^{n_{d_{\dot a}}}Z.
  \label{Fock2}
\end{eqnarray}
The constraint (\ref{C}) becomes 
\begin{eqnarray}
C=\sum_{\alpha=1}^2 n_{a_\alpha}-\sum_{\dot\alpha=1}^2 n_{b_{\dot\alpha}}+\sum_{\bar a=1}^2 n_{c_{\bar a}}-\sum_{{\dot a}=3}^4 n_{d_{\dot a}}=0.
\nonumber
\end{eqnarray}
According to this redefinition all the generators representing U(2,2$|$4) in (\ref{matrix}) 
get the central charge $C=0$.  Among them the following generators non-trivially act on $Z$, 
\begin{eqnarray}
&\ & Q^{\dot a}_{\ \beta}, \quad\quad 
{\dot Q}_{\dot\alpha \bar b},  \quad\quad 
P_{\alpha\dot\beta} , \quad\quad
R^{\dot a}_{\ \bar b},   
\nonumber
\end{eqnarray}
with the renamed indices  by  (\ref{renamee}). 
 To be explicit, they are 
\begin{eqnarray}
a^\dagger_\beta d^\dagger_{\dot a}, \quad\quad 
 b^\dagger_{\dot\alpha}c^\dagger_b, \quad\quad 
 a^\dagger_\alpha b^\dagger_{\dot\beta}, \quad\quad
c^\dagger_{\bar b} d^\dagger_{\dot a}.
\nonumber
\end{eqnarray}
Acting on $Z$ the fermionic  generators create the states 
 as shown in table 2\cite{Bei3}. They exactly correspond to  the fundamental fields of the ${\cal N}$=4 SUSY field strength multiplet in table 1. Furthermore acting on those states  $P_{\alpha\dot\beta}=a^\dagger_\alpha b^\dagger_{\dot\alpha}$ 
and $R^{\dot a}_{\ \bar b}=c^\dagger_{\bar b} d^\dagger_{\dot a}$ 
 create  SU(2)$\otimes$SU(2) excited states with the Dynkin label $[k,k]$ respectively. 
The  former excitation implies space-time derivative ${\frak D}$ of the ${\cal N}=4$ SUSY field strength  multiplet. (See table 1.) The latter excitation occurs in the representation space of  the R-symmetry SU(4).  All of these states have the central charge $C=0$, so that they are indeed in the infinite-dimensional unitary representation of PSU(2,2$|$4). The remaining generators   annihilate  $Z$. In particular the fermionic  
  ones  are   given by 
\begin{eqnarray}
&\ &Q^{\bar a}_{\ \beta}, \quad\quad {\dot Q}_{\dot\alpha \dot b}, 
\label{1/2}
\end{eqnarray}   
which  are 
\begin{eqnarray}
 a^\dagger_\beta c^{\bar a}, \quad\quad b^\dagger_{ \dot \alpha}d^{\dot b}.
\nonumber
\end{eqnarray} 
They are half of the 16 supercharges. Thus the  states in table 2 form a
half-multiplet\cite{Bei3}. They are the smallest BPS multiplet.  The vacuum $Z$ is  the highest weight vector  of the multiplet, which  is denoted by  the SU(2,2) Dynkin label [0,1,0].

\begin{table}[h]
\begin{center}
\begin{tabular}{|c|c|}  \hline
 field  &  states       \\
\hline 
$ F$  & $a^\dagger_\alpha a^\dagger_{\beta} d^\dagger_3d^\dagger_4 Z$     \\
$ \Psi$ & $a^\dagger_\alpha d^\dagger_{\dot a} Z, a^\dagger_\alpha c^\dagger_{\bar a} d^\dagger_3d^\dagger_4 Z $    \\
$ \Phi$  &  $Z$, $c^\dagger_{\bar a} d^\dagger_{\dot a}Z $, $c^\dagger_1c^\dagger_2 d^\dagger_3d^\dagger_4Z $       \\
$ \dot\Psi$ & $b^\dagger_{\dot\alpha}c^\dagger_{\bar a}Z, b^\dagger_{\dot\alpha}d^\dagger_{\dot a}c^\dagger_1c^\dagger_2 
Z$  \\
$ \dot F $ & $b^\dagger_{\dot \alpha}b^\dagger_{\dot \beta}c^\dagger_1c^\dagger_2Z$   \\
\hline
\end{tabular}
\end{center}

\vspace{-0.2cm}

\caption{Oscillator representation of the ${\cal N}=4$ SUSY field strength multiplet. }

\end{table}

 Larger BPS multiplets for the ${\cal N}=4$ SUSY theory, i.e., other ${\cal N}=4$ superconformal multiplets,  can be also constructed by generalizing the above construction. It will be done  in appendix A to complete the  argument.

\section{Matrix representation of PSU(2,2$|$4) }
\setcounter{equation}{0}

So far we have considered the unitary(oscillator) representation of U(2,2$|$4)  taking a base obtained by the tensor product  $\psi^\dagger\otimes\psi$. 
In this section we discuss a matrix representation of U(2,2$|$4) which is induced from the unitary(oscillator) representation by (\ref{state-operator1}) and (\ref{state-operator2}). To this end we put 64 generators in a 
 base  which manifests U(2,2$|$4) more faithfully than (\ref{matrix}), i.e.,  
\begin{eqnarray}
\left[
\begin{array}{c|c|c}
\hspace{0.2cm}T^\alpha_{\ \beta} \hspace{0.2cm} &\hspace{0.2cm}T^\alpha_{\ b}  \hspace{0.2cm}&\hspace{0.2cm} T^{\alpha}_{\ \dot\beta} \hspace{0.2cm}  \\
\vspace{-0.3cm}     &    &       \\
\hline
\vspace{-0.3cm}  &    &    \\
\hspace{0.2cm}T^{a}_{\ \beta} \hspace{0.2cm} & \hspace{0.2cm}T^{a}_{\ b} \hspace{0.2cm} & \hspace{0.2cm} T^{a}_{\ \dot\beta}  \hspace{0.2cm}  \\
\vspace{-0.3cm}     &      &     \\
\hline
\vspace{-0.3cm}  &   &    \\
\hspace{0.2cm}T^{\dot\alpha}_{\ \beta} \hspace{0.2cm}  &\hspace{0.2cm}T^{\dot\alpha}_{\ b}  \hspace{0.2cm} & \hspace{0.2cm} T^{\dot\alpha}_{\ \dot\beta}  \hspace{0.2cm}  
\end{array}
\right].    \label{matrixx}
\end{eqnarray}
Using an $8\times 8$ supermatrix with the index convention  
\begin{eqnarray}
\left(
\begin{array}{c|c|c}
\hspace{0.2cm}t^\gamma_{\ \delta} \hspace{0.2cm} &\hspace{0.2cm}t^\gamma_{\ d}  \hspace{0.2cm}&\hspace{0.2cm} t^{\gamma}_{\ \dot\delta} \hspace{0.2cm}  \\
\vspace{-0.3cm}     &    &       \\
\hline
\vspace{-0.3cm}  &    &    \\
\hspace{0.2cm}t^{c}_{\ \delta} \hspace{0.2cm} & \hspace{0.2cm}t^{c}_{\ d} \hspace{0.2cm} & \hspace{0.2cm} t^{c}_{\ \dot\delta}  \hspace{0.2cm}  \\
\vspace{-0.3cm}     &      &     \\
\hline
\vspace{-0.3cm}  &   &    \\
\hspace{0.2cm}t^{\dot\gamma}_{\ \delta} \hspace{0.2cm}  &\hspace{0.2cm}t^{\dot\gamma}_{\ d}  \hspace{0.2cm} & \hspace{0.2cm} t^{\dot\gamma}_{\ \dot\delta}  \hspace{0.2cm}  
\end{array}
\right),   \nonumber
\end{eqnarray}
we write  the generators as  
\begin{eqnarray}
{\scriptsize
\scriptstyle{T^\alpha_{\ \beta}}=\left(
\begin{array}{c|c|c}
\hspace{-0.1cm}\delta^\alpha_{\delta}\delta^\gamma_{\beta}\hspace{-0.1cm}  &\hspace{0.17cm}0\hspace{0.17cm}  &\hspace{0.17cm}0\hspace{0.17cm}   \\
\vspace{-0.25cm}     &    &       \\
\hline
\vspace{-0.25cm}  &    &    \\
\hspace{-0.1cm} 0\hspace{-0.1cm}   & \hspace{0.17cm}0\hspace{0.17cm}  &  \hspace{0.15cm}0\hspace{0.15cm}   \\
\vspace{-0.25cm}     &      &     \\
\hline
\vspace{-0.25cm}  &   &    \\
\hspace{-0.1cm}0\hspace{-0.1cm}   &\hspace{0.17cm}0\hspace{0.17cm}  &\hspace{0.15cm}0\hspace{0.15cm} 
\end{array}
\right)
}, \quad
{\scriptsize
\scriptstyle{T^a_{\ \beta}}=\left(
\begin{array}{c|c|c}
\hspace{0.15cm}0\hspace{0.15cm} &
\hspace{-0.0cm}\delta^a_{d}\delta^\gamma_{\beta}\hspace{-0.0cm}
&\hspace{0.15cm}0\hspace{0.15cm}\\
\vspace{-0.25cm}     &    &       \\
\hline
\vspace{-0.25cm}  &    &    \\
 \hspace{0.15cm}0\hspace{0.15cm}& \hspace{-0.1cm}0\hspace{-0.1cm}  
  &  \hspace{0.15cm}0\hspace{0.15cm}   \\
\vspace{-0.25cm}     &      &     \\
\hline
\vspace{-0.25cm}  &   &    \\
\hspace{0.15cm}0\hspace{0.15cm}
&\hspace{-0.1cm}0\hspace{-0.1cm} &\hspace{0.15cm}0\hspace{0.15cm} 
\end{array}
\right)
},\quad 
{\scriptsize
\scriptstyle{T^{\dot\alpha}_{\ \beta}}=\left(
\begin{array}{c|c|c}
\hspace{0.15cm}0\hspace{0.15cm} &\hspace{0.17cm}0\hspace{0.17cm}  &\hspace{-0.1cm}\delta^\gamma_{\beta}\delta_{\dot\delta}^{\dot\alpha}\hspace{-0.1cm}   \\
\vspace{-0.25cm}     &    &       \\
\hline
\vspace{-0.25cm}  &    &    \\
\hspace{0.15cm}0\hspace{0.15cm}
& \hspace{0.17cm}0\hspace{0.17cm}  &  \hspace{-0.1cm}0\hspace{-0.1cm}   \\
\vspace{-0.25cm}     &      &     \\
\hline
\vspace{-0.25cm}  &   &    \\
\hspace{0.15cm}0\hspace{0.15cm}
   &\hspace{0.17cm}0\hspace{0.17cm}  & \hspace{-0.1cm}0\hspace{-0.1cm}
\end{array}
\right),
}  \nonumber 
\end{eqnarray}

\vspace{-0.4cm}

\begin{eqnarray}
{\scriptsize
\scriptstyle{T^\alpha_{\ b}}=\left(
\begin{array}{c|c|c}
\hspace{-0.1cm} 0\hspace{-0.1cm}  
&\hspace{0.17cm}0\hspace{0.17cm}  &\hspace{0.15cm}0\hspace{0.15cm}   \\
\vspace{-0.25cm}     &    &       \\
\hline
\vspace{-0.25cm}  &    &    \\
\hspace{-0.1cm}\delta^\alpha_{\delta}\delta^c_{ b}\hspace{-0.1cm} 
&  \hspace{0.17cm}0\hspace{0.17cm}  &  \hspace{0.15cm}0\hspace{0.15cm} \\
\vspace{-0.25cm}     &      &     \\
\hline
\vspace{-0.25cm}  &   &    \\
\hspace{-0.1cm}0\hspace{-0.1cm}   &\hspace{0.17cm}0\hspace{0.17cm}  &\hspace{0.15cm}0\hspace{0.15cm} 
\end{array}
\right)
}, \quad
{\scriptsize
\scriptstyle{T^a_{\ b}}=\left(
\begin{array}{c|c|c}
\hspace{0.15cm}0\hspace{0.15cm} &  \hspace{-0.1cm}0\hspace{-0.1cm}
&\hspace{0.15cm}0\hspace{0.15cm}\\
\vspace{-0.25cm}     &    &       \\
\hline
\vspace{-0.25cm}  &    &    \\
\hspace{0.15cm}0\hspace{0.15cm}
 &\hspace{-0.0cm}\delta^a_{d}\delta^c_{ b}\hspace{-0.0cm}
 &  \hspace{0.15cm}0\hspace{0.15cm}   \\
\vspace{-0.25cm}     &      &     \\
\hline
\vspace{-0.25cm}  &   &    \\
\hspace{0.15cm}0\hspace{0.15cm}
  &\hspace{-0.1cm}0\hspace{-0.1cm}  &\hspace{0.15cm}0\hspace{0.15cm} 
\end{array}
\right)
},\quad 
{\scriptsize
\scriptstyle{T^{\dot\alpha}_{\ b}}=\left(
\begin{array}{c|c|c}
\hspace{0.15cm}0\hspace{0.15cm} &\hspace{0.17cm}0\hspace{0.17cm}  &\hspace{-0.1cm}0\hspace{-0.1cm}  \\
\vspace{-0.25cm}     &    &       \\
\hline
\vspace{-0.25cm}  &    &    \\
\hspace{0.15cm}0\hspace{0.15cm}  
& \hspace{0.17cm}0\hspace{0.17cm}  & 
\hspace{-0.1cm}\delta^c_{b}\delta_{\dot\delta}^{\dot\alpha}\hspace{-0.1cm} 
   \\
\vspace{-0.25cm}     &      &     \\
\hline
\vspace{-0.25cm}  &   &    \\
\hspace{0.15cm}0\hspace{0.15cm}   &\hspace{0.17cm}0\hspace{0.17cm}  &\hspace{-0.1cm}0\hspace{-0.1cm} 
\end{array}
\right),
}   \label{matrixrep} 
\end{eqnarray}

\vspace{-0.4cm}

\begin{eqnarray}
{\scriptsize
\scriptstyle{T^{\alpha}_{\ \dot\beta}}=-\left(
\begin{array}{c|c|c}
\hspace{-0.1cm}0\hspace{-0.1cm} 
 &\hspace{0.17cm}0\hspace{0.17cm}  &\hspace{0.15cm}0\hspace{0.15cm}   \\
\vspace{-0.25cm}     &    &       \\
\hline
\vspace{-0.25cm}  &    &    \\
\hspace{-0.1cm} 0\hspace{-0.1cm}   & \hspace{0.17cm}0\hspace{0.17cm}  &  \hspace{0.15cm}0\hspace{0.15cm}   \\
\vspace{-0.25cm}     &      &     \\
\hline
\vspace{-0.25cm}  &   &    \\
\hspace{-0.1cm}\delta^\alpha_{\delta}\delta_{\dot\beta}^{\dot\gamma}\hspace{-0.1cm} 
 &\hspace{0.17cm}0\hspace{0.17cm}  &\hspace{0.15cm}0\hspace{0.15cm} 
\end{array}
\right)
}, \quad
{\scriptsize
\scriptstyle{T^{a}_{\ \dot\beta}}=-\left(
\begin{array}{c|c|c}
\hspace{0.15cm}0\hspace{0.15cm} &
\hspace{-0.1cm}0\hspace{-0.1cm}
&\hspace{0.15cm}0\hspace{0.15cm}\\
\vspace{-0.25cm}     &    &       \\
\hline
\vspace{-0.25cm}  &    &    \\
\hspace{0.15cm}0\hspace{0.15cm} 
& \hspace{-0.1cm}0\hspace{-0.1cm} 
&  \hspace{0.15cm}0\hspace{0.15cm}   \\
\vspace{-0.25cm}     &      &     \\
\hline
\vspace{-0.25cm}  &   &    \\
\hspace{0.15cm}0\hspace{0.15cm}
&\hspace{-0.0cm}\delta^{\dot\gamma}_{\dot\beta}\delta^a_{ d}\hspace{-0.0cm}
&\hspace{0.15cm}0\hspace{0.15cm} 
\end{array}
\right)
},\quad 
{\scriptsize
\scriptstyle{T^{\dot\alpha}_{\ \dot\beta}}=-\left(
\begin{array}{c|c|c}
\hspace{0.15cm}0\hspace{0.15cm} &\hspace{0.17cm}0\hspace{0.17cm}  &
\hspace{-0.1cm}0\hspace{-0.1cm}
   \\
\vspace{-0.25cm}     &    &       \\
\hline
\vspace{-0.25cm}  &    &    \\
\hspace{0.15cm}0\hspace{0.15cm}
 & \hspace{0.17cm}0\hspace{0.17cm}  & \hspace{-0.1cm}0\hspace{-0.1cm}
   \\
\vspace{-0.25cm}     &      &     \\
\hline
\vspace{-0.25cm}  &   &    \\
\hspace{0.15cm}0\hspace{0.15cm}
&\hspace{0.17cm}0\hspace{0.17cm}  &
\hspace{-0.1cm}\delta^{\dot\alpha}_{\dot\delta}\delta^{\dot\gamma}_{\dot\beta}\hspace{-0.1cm}
\end{array}
\right).
}  \nonumber 
\end{eqnarray}
Here keep in mind the {\it minus} sign in the last line which accounts for non-compactness of U(2,2$|$4). Bosonic generators in the diagonal blocks of (\ref{matrixx}) form the Lie-algebra of U(2,2)$\otimes$U(4) 
\begin{eqnarray}
\ [T^\alpha_{\ \beta}, T^\gamma_{\ \delta}]&=& -\delta_\beta^\gamma T^\alpha_{\ \delta}+\delta^\alpha_\delta T^\gamma_{\ \beta}, \quad\quad 
[T^{\dot\alpha}_{\ \dot\beta}, T^{\dot\gamma}_{\ \dot\delta}]= \delta_{\dot\beta}^{\dot\gamma} T^{\dot\alpha}_{\ \dot\delta}-\delta^{\dot\alpha}_{\dot\delta} T^\gamma_{\ \beta},  
  \nonumber\\
&\ & \hspace{1cm}[T^a_{\ b}, T^c_{\ d}]=-\delta_b^c T^a_{\ d}+\delta^a_d T^c_{\ d},  \label{bosonalg2} \\
&\ & \hspace{1cm}[T^\alpha_{\ \dot\beta}, T^{\dot\gamma}_{\ \delta}]= 
\delta^{\dot\gamma}_{\dot\beta}T^\alpha_{\ \delta}  
-\delta^\alpha_{ \delta}T^{\dot\gamma}_{\ \dot\beta}.   \nonumber
\end{eqnarray}
Anti-commuting  fermionic generators  in the off-diagonal blocks with each other yields 
\begin{eqnarray}
\{T^\alpha_{\ b},T^a_{\ \beta}\}&=&\delta^\alpha_\beta T^a_{\ b}+\delta^a_b T^\alpha_{\ \beta},\quad\quad 
\{T^a_{\ \dot\beta},T^{\dot\alpha}_{\  b}\}=-\delta^a_b T^{\dot\alpha}_{\ \dot\beta} + \delta^{\dot\alpha}_{\dot\beta}T^a_{\ b},   \nonumber\\
\{T^\alpha_{\ b},T^a_{\ \dot\beta}\} &=& \delta^a_b T^\alpha_{\ \dot\beta},\quad\quad\quad  \hspace{1cm}
\{T^a_{\ \beta},T^{\dot\alpha}_{\  b}\}= \delta^a_b T^{\dot\alpha}_{\ \beta}.
 \label{fermionalg2}
\end{eqnarray}
Commuting these fermionic generators  with bosonic generators  yields 
\begin{eqnarray}
[T^\alpha_{\ b},T^\gamma_{\ \delta}]&=& \delta^\alpha_\delta T^\gamma_{\ b}, \quad\quad \ \  
[T^\alpha_{\ b},T^c_{\ d}]=-\delta^c_b T^\alpha_{\ d},  
\quad\quad [T^\alpha_{\ b},T^{\dot\gamma}_{\ \dot\delta}]=0,
\nonumber\\
\ [T^a_{\ \beta},T^\gamma_{\ \delta}]&=& -\delta^\gamma_\beta T^a_{\ \delta}, \quad\quad [T^a_{\ \beta},T^c_{\ d}]=\delta^a_d T^c_{\ \beta},  
\quad\quad \ [T^a_{\ \beta},T^{\dot\gamma}_{\ \dot\delta}]=0, 
 \nonumber\\
\ [T^a_{\ \dot\beta},T^{\gamma}_{\ \delta}]&=& 0, \quad\quad \quad\quad\
[T^a_{\ \dot\beta},T^c_{\ d}]=\delta^a_d T^c_{\ \dot\beta}, 
\quad\quad\  [T^a_{\ \dot\beta},T^{\dot\gamma}_{\ \dot\delta}]=\delta^{\dot\gamma}_{\dot\beta} T^a_{\ \dot\delta}, \label{fermiboson}\\
\ [T^{\dot\alpha}_{\ b},T^{\gamma}_{\ \delta}]&=& 0,  \quad\quad \quad\quad\
 [T^{\dot\alpha}_{\ b},T^c_{\ d}]=-\delta^c_b T^{\dot\alpha}_{\ d}, \quad\quad
[T^{\dot\alpha}_{\ b},T^{\dot\gamma}_{\ \dot\delta}]= -\delta^{\dot\alpha}_{\ \dot\delta}T^{\dot\gamma}_{\ b}.
 \nonumber 
\end{eqnarray}
while commuting them  with bosonic generators  
\begin{eqnarray}
[T^\alpha_{\ b},T^\gamma_{\ \dot\delta}]&=& 0, \quad\quad\quad\quad\ \ [T^\alpha_{\ b},T^{\dot\gamma}_{\ \delta}]=\delta^\alpha_\delta T^{\dot\gamma}_{\ b},  \nonumber  \\
\ [T^a_{\ \beta},T^\gamma_{\ \dot\delta}]&=& -\delta^\gamma_\beta T^a_{\ \dot\delta}, \quad\quad [T^a_{\ \beta},T^{\dot\gamma}_{\ \delta}]=0,    \nonumber \\
\ [T^a_{\ \dot\beta},T^{\gamma}_{\ \dot\delta}]&=& 0,\quad\quad \quad\quad  
[T^a_{\ \dot\beta},T^{\dot\gamma}_{\ \delta}   ]=\delta^{\dot\gamma}_{\dot\beta} T^a_{\ \delta},  \label{fermiboson3}\\
\ [T^{\dot\alpha}_{\ b},T^{\gamma}_{\ \dot\delta}]&=& -\delta^{\dot\alpha}_{\ \dot\delta}T^{\gamma}_{\ b}, \quad\quad 
 [T^{\dot\alpha}_{\ b},T^{\dot\gamma}_{\ \delta}]=0.   \nonumber 
\end{eqnarray}
All other (anti-)commutation relations are vanishing.  The diagonal blocks
 contain the generators of the subgroup SU(2)$\otimes$SU(2)$\otimes$SU(4) given by 
\begin{eqnarray}
 {\cal L}^\alpha_{\ \beta}=T^\alpha_{\ \beta}-{1\over 2}\delta^\alpha_{\beta}T^\gamma_{\ \gamma},\quad\quad {\dot{\cal L}}^{\dot\alpha}_{\ \dot\beta}=T^{\dot\alpha}_{\ \dot\beta}-{1\over 2}\delta^{\dot\alpha}_{\dot\beta}T^{\dot\gamma}_{\ \dot\gamma},\quad\quad {\cal R}^a_{\ b}=T^a_{\ b}-{1\over 4}\delta^a_b T^c_{\ c},  \label{traceless}
\end{eqnarray}
and three U(1) generators defined by 
\begin{eqnarray}
D&=& 
{1\over 2}\left(
\begin{array}{c|c|c}
\hspace{0.1cm}\delta^\gamma_{\ \delta} \hspace{0.1cm} &\hspace{0.25cm}0 \hspace{0.25cm}&\hspace{-0.05cm} 0 \hspace{-0.05cm}  \\
\vspace{-0.3cm}     &    &       \\
\hline
\vspace{-0.3cm}  &    &    \\
\hspace{0.1cm}0\hspace{0.1cm} & \hspace{0.25cm}0 \hspace{0.25cm} & \hspace{-0.05cm} 0 \hspace{-0.05cm}  \\
\vspace{-0.3cm}     &      &     \\
\hline
\vspace{-0.3cm}  &   &    \\
\hspace{0.1cm}0 \hspace{0.1cm}  &\hspace{0.25cm} 0 \hspace{0.25cm} & \hspace{-0.05cm}- \delta^{\dot\gamma}_{\ \dot\delta}  \hspace{-0.05cm}  
\end{array}
\right)={1\over 2}( T^\alpha_{\ \alpha} +T^{\dot\alpha}_{\ \dot\alpha}),                      \nonumber \\
-C&=& 
{1\over 2}\left(
\begin{array}{c|c|c}
\hspace{0.1cm}\delta^\gamma_{\ \delta} \hspace{0.11cm} &\hspace{0.11cm}0 \hspace{0.1cm}&\hspace{0.1cm} 0 \hspace{0.1cm}  \\
\vspace{-0.3cm}     &    &       \\
\hline
\vspace{-0.3cm}  &    &    \\
\hspace{0.1cm}0\hspace{0.1cm} & \hspace{0.11cm}\delta^c_{\ d} \hspace{0.11cm} & \hspace{0.1cm} 0 \hspace{0.1cm}  \\
\vspace{-0.3cm}     &      &     \\
\hline
\vspace{-0.3cm}  &   &    \\
\hspace{0.1cm}0 \hspace{0.1cm}  &\hspace{0.11cm} 0 \hspace{0.11cm} & \hspace{0.1cm} \delta^{\dot\gamma}_{\ \dot\delta}  \hspace{0.1cm}  
\end{array}\right) 
={1\over 2}(T^\alpha_{\ \alpha} -T^{\dot\alpha}_{\ \dot\alpha}+T^a_{\ a}),   \label{B}     \\
B&=& 
{1\over 2}\left(
\begin{array}{c|c|c}
\hspace{0.1cm}\delta^\gamma_{\ \delta} \hspace{0.1cm} &\hspace{0.25cm}0 \hspace{0.25cm}&\hspace{0.1cm} 0 \hspace{0.1cm}  \\
\vspace{-0.3cm}     &    &       \\
\hline
\vspace{-0.3cm}  &    &    \\
\hspace{0.1cm}0\hspace{0.1cm} & \hspace{0.25cm}0 \hspace{0.25cm} & \hspace{0.1cm} 0 \hspace{0.1cm}  \\
\vspace{-0.3cm}     &      &     \\
\hline
\vspace{-0.3cm}  &   &    \\
\hspace{0.1cm}0 \hspace{0.1cm}  &\hspace{0.25cm} 0 \hspace{0.25cm} & \hspace{0.1cm} \delta^{\dot\gamma}_{\ \dot\delta}  \hspace{0.1cm}  
\end{array}\right) ={1\over 2}
(T^\alpha_{\ \alpha} -T^{\dot\alpha}_{\ \dot\alpha}).  \nonumber       
\end{eqnarray}
They are identical to the one given in the unitary(oscillator) representation (\ref{U(1)}).
U(2,2$|$4) becomes SU(2,2$|$4) or PU(2,2$|$4) when the U(1) generators are constrained by $B=0$ or $C=0$ respectively. When imposed both constraints, it becomes PSU(2,2$|$4). 

Using the generators defined by (\ref{traceless}) we rewrite the algebrae (\ref{bosonalg2})$\sim$(\ref{fermiboson}).
 The first three algebrae in (\ref{bosonalg2}) remain in the same form 
\begin{eqnarray}
\ [{\cal L}^\alpha_{\ \beta}, {\cal L}^\gamma_{\ \delta}]&=& -\delta_\beta^\gamma {\cal L}^\alpha_{\ \delta}+\delta^\alpha_\delta {\cal L}^\gamma_{\ \beta}, \quad\quad 
[{\dot{\cal L}}^{\dot\alpha}_{\ \dot\beta}, {\dot{\cal L}}^{\dot\gamma}_{\ \dot\delta}]= \delta_{\dot\beta}^{\dot\gamma} {\dot{\cal L}}^{\dot\alpha}_{\ \dot\delta}-\delta^{\dot\alpha}_{\dot\delta} {\dot{\cal L}}^{\dot\gamma}_{\ \dot\beta},  
 \nonumber \\
&\ & \hspace{1cm}[{\cal R}^a_{\ b}, {\cal R}^c_{\ d}]=-\delta_b^c {\cal R}^a_{\ d}+\delta^a_d {\cal R}^c_{\ d}.    \label{L}
\end{eqnarray}
Other algebrae in (\ref{bosonalg2})$\sim$(\ref{fermiboson}) 
 also do not change  significantly the forms, except for the last algebra in (\ref{bosonalg2}) and the first two  in  (\ref{fermionalg2}). Those are found to be 
\begin{eqnarray}
[T^\alpha_{\ \dot\beta}, T^{\dot\gamma}_{\ \delta}]&=& 
\delta^{\dot\gamma}_{\dot\beta}{\cal L}^\alpha_{\ \delta} 
+\delta^\alpha_{ \delta}{\dot{\cal L}}^{\dot\gamma}_{\ \dot\beta} 
+\delta^{\dot\gamma}_{\dot\beta}\delta^\alpha_{\ \delta}D,     \nonumber\\
\{T^\alpha_{\ b},T^a_{\ \beta}\}&=&\delta^\alpha_\beta {\cal R}^a_{\ b}+\delta^a_b{\cal L}^\alpha_{\ \beta}+{1\over 2}\delta^\alpha_\beta\delta^a_b (D-C),     \label{except}\\
\{T^a_{\ \dot\beta},T^{\dot\alpha}_{\  b}\}&=&\delta^a_b {\dot{\cal L}}^{\dot\alpha}_{\ \dot\beta} - \delta^{\dot\alpha}_{\dot\beta}{\cal R}^a_{\ b}+{1\over 2}\delta^a_b\delta^{\dot\alpha}_{\dot\beta}(D+C).   \nonumber
\end{eqnarray}
The quadratic Casimir is given by 
\begin{eqnarray}
 {\cal L}^\alpha_{\ \beta} {\cal L}^\beta_{\ \alpha}-{\cal R}^a_{\ b}{\cal R}^b_{\ a}+{\dot{\cal L}}^{\dot\alpha}_{\ \dot\beta} {\dot{\cal L}}^{\dot\beta}_{\ \dot\alpha}+D^2 -\{T^{\alpha}_{\ \dot\beta},T^{\dot\beta}_{\ \alpha}\}-
[T^a_{\ \beta},T^\beta_{\ a} ]-[T^{\dot\beta}_{\ a},T^{a}_{\ \dot\beta}].  \label{Casimir}
\end{eqnarray}

 Now we compare the algebrae (\ref{bosonalg2})$\sim$(\ref{fermiboson3}) with 
 (\ref{L0})$\sim$(\ref{L0+}) in the unitary(oscillator) representation. We find them to be equivalent by redefining the generators as
\begin{eqnarray}
&\ & \epsilon^{\dot\alpha\dot\delta}\epsilon_{\dot\beta\dot\gamma} {\dot{\cal L}}^{\dot\gamma}_{\ \dot\delta}={\dot L}^{\dot\alpha}_{\ \dot\beta}=-{\dot{\cal L}}^{\dot\alpha}_{\ \dot\beta}, \quad\quad
\epsilon^{\dot\beta\dot\delta}T^\alpha_{\ \dot\delta}=K^{\alpha\dot\beta},
\quad\quad \epsilon_{\dot\alpha\dot\gamma}T^{\dot\gamma}_{\ \beta}=P_{\dot\alpha\beta},    \nonumber\\
&\ & \hspace{1.5cm} \epsilon^{\dot\beta\dot\delta}T^a_{\ \dot\delta}={\dot S}^{a\dot\beta},
\quad\quad \epsilon_{\dot\alpha\dot\gamma}T^{\dot\gamma}_{\ b}={\dot Q}_{\dot\alpha b}, \nonumber 
\end{eqnarray}
\vspace{-0.5cm}
$$
{\cal L}^\alpha_{\ \beta}=L^\alpha_{\ \beta}, {\cal R}^a_{\ b}=R^a_{\ b},\quad\quad {\cal S}^\alpha_{\ b}=T^\alpha_{\ b},\quad\quad T^a_{\ \beta} =Q^a_{\ \beta},\quad\quad{\cal D}=D. 
\vspace{0.2cm}
$$
The redefinition does not change the form of the quadratic Casimir (\ref{Casimir}). It coincides with the quadratic Casimir (\ref{Casimir0}), given in the unitary(oscillator) representation. But the redefintion changes the sign of the algebrae linearly containing ${\dot{\cal L}}^{\dot\alpha}_{\ \dot\beta}$ in (\ref{bosonalg2})$\sim$(\ref{fermiboson3}). For instance, the second one in (\ref{L}) becomes that of (\ref{L0}).

We compare also   the algebra in  (\ref{bosonalg2})$\sim$(\ref{fermiboson3})  with those of U(4$|$4) and U(8). If the matrices (\ref{matrixrep})  get all entries with {\it plus} sign, i.e., 
 $(T^A_{\ B})^C_{\ D}=\delta^A_D\delta^C_B$, 
  they become the generators of U(4$|$4). They  satisfy the Lie-algebrae 
(\ref{bosonalg2})$\sim$(\ref{fermiboson3})  where $T^\alpha_{\ \dot\beta}, T^a_{\ \dot\beta},  T^{\dot\alpha}_{\ \dot\beta}$ get the sign changed.  Accordingly the quadratic Casimir (\ref{Casimir}) changes the form as
\begin{eqnarray}
&\ &\sum_{B=\beta,\dot\beta,b}T^\alpha_{\ B}T^B_{\ \alpha}+
\sum_{B=\beta,\dot\beta,b}T^{\dot\alpha}_{\ B}T^B_{\ \dot\alpha}
-\sum_{B=\beta,\dot\beta,b}T^a_{\ B}T^B_{\ a}=(-1)^{g(A)}T^A_{\ B}T^B_{\ A}.  \nonumber
\end{eqnarray} 
Here we have assigned the grading $g(A)$ to the index $A$ in such a way 
 $g(A)=1$ for a  fermionic index and otherwise $g(A)=0$. So $T^A_{\ B}$ has the grading $g(A)g(B)$. 
 If we do not assign the grading, $T^A_{\ B}$ satisfies  the Lie-algebra of U(8)
$$
[T^A_{\ B},T^C_{\ D}]= -\delta^C_B T^A_{\ D}+\delta^A_DT^C_{\ B}.
$$
being defined as  $(T^A_{\ B})^C_{\ D}=\delta^A_D\delta^C_B$, 
The the quadratic Casimir of U(8) is simply
\begin{eqnarray}
\sum_{B=\beta,\dot\beta,b}T^\alpha_{\ B}T^B_{\ \alpha}+
\sum_{B=\beta,\dot b}T^{\dot\alpha}_{\ B}T^B_{\ \dot\alpha}
+\sum_{B=\beta,\dot\beta,b}T^a_{\ B}T^B_{\ a}=T^A_{\ B}T^B_{\ A}.  \nonumber
\end{eqnarray}
Or we had better formulate   the superalgebrae of  U(2,2$|$4) and U(4$|$4) in a converse way, i.e.,  starting with this  form of the algebra of U(8) instead of the graded form  of (\ref{bosonalg2})$\sim$(\ref{fermiboson3}).

\section{Non-linear realization of PSU(2,2$|$4)} 
\setcounter{equation}{0}

Both the  unitary(oscillator) representation and  the matrix one  allow  linear realization of PSU(2,2$|$4)  only as a subgroup of its centrally extended group SU(2,2$|$4). It can been seen from the respective algebrae (\ref{L+-}) and (\ref{except}). PSU(2,2$|$4) is  a simple group so that we do not need  the central extension at the algebraic level. In this section 
we want to discuss a purely algebraic method to non-linearly  realize PSU(2,2$|$4), which does not rely on the explicit representations and is consequently free from the central charge of SU(2,2$|$4). 
To this end we begin by writing the Lie-algebra of PSU(2,2$|$4) in a common form by which we can freely change the  unitary(oscillator) representation to the matrix one and {\it vice versa}. Then using that algebra 
 we give general accounts of non-linear realization of  PSU(2,2$|$4) on the coset space PSU(2,2$|$4)/H, without being bothered by the specifics of a chosen subgroup H. 
We discuss afterwards the case where H is  SO(1,4)$\otimes$SO(5), which is the main concern in this paper. 

\subsection{Algebraic method of non-linear realization}

Let us  put the generators of PU(2,2$|$4) in a row  and denote them   by $\{T^{\Xi}\}$.
 That is, 62 generators in the unitary(oscillator) representation, discussed in section 3,  are 
  denoted by  
\begin{eqnarray}
\{T^{\Xi}\}= 
\{L^\alpha_{\ \beta},{\dot L}^{\dot\alpha}_{\ \dot\beta},R^a_{\ b}, D,P_{\dot\alpha\beta},K^{\beta\dot\alpha},S^\alpha_{\ b},Q^a_{\ \alpha},{\dot S}^{a\dot\alpha},{\dot Q}_{\dot\alpha a}\}, \label{set1}
\end{eqnarray}
while the corresponding generators in the matrix representation, discussed in section 4, by 
\begin{eqnarray}
\{T^{\Xi}\}=
\{{\cal L}^\alpha_{\ \beta},{\dot{\cal L}}^{\dot\alpha}_{\ \dot\beta},{\cal R}^a_{\ b}, {\cal D},{\cal P}_{\dot\alpha\beta},{\cal K}^{\beta\dot\alpha},{\cal S}^\alpha_{\ b},{\cal Q}^a_{\ \alpha},{\dot {\cal S}}^{a\dot\alpha},{\dot {\cal Q}}_{\dot\alpha a}\}.  \label{set2}
\end{eqnarray}
Using either set of these 62 generators  we represent  PSU(2,2$|$4)
in a common form  as
\begin{eqnarray}
e^{iM^\Xi T^\Xi}  \in {\rm PSU(2,2}|{\rm 4)},
 \label{standard}
\end{eqnarray}
in which $M^\Xi$ are 62 elements of the supermatrix $M$ given in (\ref{M}) 
\begin{eqnarray}
\{M^\Xi\}&=&\Big\{V^\alpha_{\ \beta}-{1\over 2}\delta^\alpha_\beta, Z^{\dot\alpha}_{\ \dot\beta}-{1\over 2}\delta^{\dot\alpha}_{\dot\beta},W^a_{\ b}-{1\over 4}\delta^a_b, \nonumber\\
&\ &\hspace{4cm} D,-X^\dagger_{\dot\alpha\beta},X^{\beta\dot\alpha},\theta^{\alpha}_{\  b},\theta^{\dagger a}_{\ \ \alpha},\epsilon^{a\dot\alpha},-\epsilon^\dagger_{\dot\alpha a}\Big\}. 
\label{elements}  \nonumber
\end{eqnarray}  
We find explicit  forms of $M^\Xi T^\Xi$ for the respective representations, expanding 
$\bar\psi M\psi$ and $M$ in terms of the generators (\ref{set1}) and (\ref{set2}). The expansion of the former  reads
\begin{eqnarray}
\bar\psi M\psi&\equiv& \sum_{B=\beta,\dot\beta,b}\psi^\dagger_\alpha(\gamma M)^\alpha_{\ B}\psi^B  
 \sum_{B=\beta,\dot\beta,b}\psi^\dagger_{\dot\alpha}(\gamma M)^{\dot\alpha}_{\ B}\psi^B   
+ \sum_{B=\beta,\dot\beta,b}\psi^\dagger_a(\gamma M)^a_{\ B}\psi^B
\nonumber\\
&=& \Big[V^\alpha_{\ \beta} L^\beta_{\ \alpha}
  +X^{\alpha\dot\beta}P_{\dot\beta\alpha} +\theta^{\alpha}_{\ \ a}Q^a_{\ \alpha}\Big] 
+\Big[(-Z^{\dot\alpha}_{\ \dot\beta} L^{\dot\beta}_{\ \dot\alpha}+Z^{\dot\gamma}_{\ \dot\gamma}) +
X^\dagger_{\dot\alpha\beta}K^{\beta\dot\alpha} +\epsilon^\dagger_{\dot\alpha a}{\dot S}^{a\dot\alpha}\Big]   \nonumber \\
&+& \Big[W^a_{\ b}R^b_{\ a}-\theta^{\dagger a}_{\ \alpha}S^\alpha_{\ a}
-\epsilon^{a\dot\alpha}{\dot Q}_{\dot\alpha a}\Big]
 +{1\over 2}(V^\gamma_{\ \ \gamma}-Z^{\dot\gamma}_{\ \dot\gamma}) D  \nonumber\\
&+&{1\over 2}(V^\gamma_{\ \ \gamma}+Z^{\dot\gamma}_{\ \dot\gamma}) B
-{1\over 2}W^c_{\ c}(B+C) 
   \nonumber\\
&\equiv &   M^{\Xi}T^\Xi+{1\over 2}(V^\gamma_{\ \ \gamma}+Z^{\dot\gamma}_{\ \dot\gamma}) B
-{1\over 2}W^c_{\ c}(B+C),        \label{oscirep} 
\end{eqnarray}
 by using the commutation relations (\ref{commutation}) and the generators defined by (\ref{LLR}) and (\ref{U(1)}). 
On the other hand the expansion of the latter reads
\begin{eqnarray}
M&=& \sum_{B=\beta,\dot\beta,b}M^\alpha_{\ B}\cdot\gamma T^B_{\ \alpha}+
\sum_{B=\beta,\dot\beta,b}M^{\dot\alpha}_{\ B}\cdot\gamma T^B_{\ \dot\alpha} 
+\sum_{B=\beta,\dot\beta,b}M^a_{\ B}\cdot\gamma T^B_{\ a}  \nonumber\\
&=& \Big[V^\alpha_{\ \beta}{\cal L}^\beta_{\ \alpha}+X^{\alpha\dot\beta}{\cal P}_{\dot\beta\alpha} +\theta^{\alpha}_{\ \ b}{\cal Q}^a_{\ \alpha}\Big] 
+\Big[Z^{\dot\alpha}_{\ \dot\beta}(- {\dot{\cal L}}^{\dot\beta}_{\ \dot\alpha}) -
X^\dagger_{\dot\alpha\beta}(-{\cal K}^{\beta\dot\alpha}) -\epsilon^\dagger_{\dot\alpha a}(-{\dot {\cal S}}^{a\dot\alpha})\Big]  \nonumber\\
 &+& \Big[W^a_{\ b}{\cal R}^b_{\ a}+\theta^{\dagger a}_{\ \alpha}{\cal S}^\alpha_{\ b}
+\epsilon^{a\dot\alpha}{\dot {\cal Q}}_{\dot\alpha a}\Big]
 +{1\over 2}(V^\gamma_{\ \ \gamma}-Z^{\dot\gamma}_{\ \dot\gamma}) {\cal D} \nonumber\\
&+&  {1\over 2}(V^\gamma_{\ \ \gamma}+Z^{\dot\gamma}_{\ \dot\gamma}) {B} 
-{1\over 2}W^c_{\ c}(B+C)   \nonumber\\
&\equiv& M^{\Xi}T^\Xi+{1\over 2}(V^\gamma_{\ \ \gamma}+Z^{\dot\gamma}_{\ \dot\gamma}) {B} 
-{1\over 2}W^c_{\ c}(B+C),
                  \label{Mrep}
\end{eqnarray}
by using the generators defined by  (\ref{traceless}) and (\ref{B}) and noting 
  $(\gamma T^A_{\ B})^C_{\ D}=\delta^A_D\delta^C_B$. A sign  difference in the third square brackets $[\cdots]$ of the the respective expansions  (\ref{oscirep}) and (\ref{Mrep}) does not indicate anything wrong. This is due to the different prescription in grading  $T^\Xi$ in both representations. 
In (\ref{oscirep}) we have employed the prescription 
\begin{eqnarray}
T^\Xi M^\Phi=(-1)^{g(\Xi)g(\Phi)}M^\Phi T^\Xi.  \label{Pres1}
\end{eqnarray}
Here the grading of $T^\Xi$ is the same as $M^\Xi$, i.e., $g(\Xi)=g(A)g(B)$ when  $\{T^\Xi\}$ is put in the tensor form $\{\psi^{\dagger A}\psi^B\}$ 
as (\ref{matrix}). 
Hence $\bar\psi M\psi$ is a bosonic operator acting on the Fock space (\ref{Fock2}). On the other hand, in  (\ref{Mrep}) we have employed the prescription 
\begin{eqnarray}
T^\Xi M^\Phi=M^\Phi T^\Xi,    \label{Pres2}
\end{eqnarray}
assigning no grading to $T^\Xi$. This is also reasonable because 
the generators (\ref{set2})  consist of bosonic elements as (\ref{matrixrep}) and commute any element of $M$. The reader may see appendix B for more arguments on these prescriptions. 

It is the fact  that $\exp(i\bar\psi M\psi)$ and $\exp(iM)$ with (\ref{oscirep}) and (\ref{Mrep}) are related  by the operator-state relations (\ref{state-operator1}) and (\ref{state-operator2}). Note that owing to these relations  the multiplication 
$ 
\exp(i\bar\psi M_1\psi)
\times$ $ \exp(i\bar\psi M_2\psi) 
$ in the Fock space  induces that of supermatrices as $M_1 M_2$. 
Thus  we are now in a position to discuss the coset space PSU(2,2$|$4)/H in either of the representations. By using the common  form of the representation (\ref{standard})  we can freely change  one representation to another in the following discussion. 
Decompose  the  generators of PSU(2,2$|$4) , given by either (\ref{set1}) or (\ref{set2}), under a subgroup H as
\begin{eqnarray}
\{T^\Xi\}=\{T^i,H^I\},  \label{decompo}
\end{eqnarray}
in which $H^I$ are generators of H, while $T^i$ coset ones. Then 
 we consider a coset element 
\begin{eqnarray}
e^{i\phi^1 T^1+i\phi^2 T^2+\cdots}\equiv e^{i\phi\cdot T}. \label{G/H}
\end{eqnarray}
Here $\phi^1,\phi^2,\cdots$, are coordinates reparametrizing the coset space, denoted by $\phi^{\bar i}$. (We keep the index $i$ for indicating  a vector component  in the tangent frame as (\ref{decompo}) .) Of course they have the same grading as the coset part of $M^\Xi$, i.e., $g(\bar i)=g(\Xi)|_{\Xi=i}$. For left multiplication of an element $e^{iM^{\Xi}T^\Xi}\in {\rm G}$, (\ref{standard}),  the coset element changes 
 as
\begin{eqnarray}
e^{iM^{\Xi}T^\Xi}e^{\phi\cdot T}=e^{\phi'(\phi)\cdot T}e^{i\rho(\phi,M)},
\label{nonlinear}
\end{eqnarray}
with an appropriate compensator $e^{i\rho(\phi,M)}$.  
This defines a transformation of the coordinates $\phi^{\bar i}\rightarrow \phi'^{\bar i}(\phi)$. When $M^{\Xi}$ are infinitesimally small, this relation defines the Killing vectors $R^{\Xi {\bar i}}$   as
\begin{eqnarray}
\delta \phi^{\bar i}=\phi'^{\bar i}(\phi)-\phi^{\bar i}\equiv M^{\Xi}R^{\Xi {\bar i}}\equiv M^\Xi\delta^\Xi\phi^{\bar i}.  \label{Killing}
\end{eqnarray}
They satisfy the Lie-algebra of PSU(2,2$|$4) 
\begin{eqnarray}
R^{\Xi {\bar i}}{\partial \over \partial \phi^{\bar i}}R^{\Phi {\bar j}}-(-1)^{g(\Xi)g(\Phi)}
R^{\Phi {\bar i}}{\partial \over \partial \phi^{\bar i}}R^{\Xi {\bar j}}=f^{\Xi\Phi\Sigma}R^{\Sigma {\bar j}},
\label{Liealgebra}
\end{eqnarray}
with the structure constants $f^{\Xi\Phi\Sigma}$ of PSU(2,2$|$4).

We would like to make important comments on the above algebraic construction. First of all,  the construction does not need  any representation at all, although we have proceeded the arguments having the unitary(oscillator)  representation or the matrix representation in mind. That is, the above machinery to construct the Killing vectors $R^{\Xi {\bar i}}$ works at the algebraic level, once given the Lie-algebra of the generators  $T^\Xi$. We give a demonstration for this in appendix B. Hence the forms of the  Killing vectors $R^{\Xi {\bar i}}$ are the same if two representations take the same form of the Lie-algebra,  like the unitary(oscillator) representation (\ref{set1}) and the matrix one (\ref{set2}). 
Moreover the  Killing vectors $R^{\Xi {\bar i}}$ are  free from any extra U(1) factor of the central charge $C$,  since the calculation is purely algebraic. On the contrary,  if the construction is done  by using  the unitary(oscillator)  representation or the matrix one, in (\ref{nonlinear}) the compensator acquires  an extra U(1) factor as
$$
e^{\rho(\phi,M)}= e^{\rho(\phi,M)^I H^I + c(\phi,M)C} \in H\otimes U(1),
$$
 even though $e^{iM^\Xi T^\Xi}$ does not have it. Here $\rho(\phi,M)^I$ and $c(\phi,M)$ are appropriate functions of $\phi^{\bar i}$. 
This is due to the fact that the Lie-algebra of PSU(2,2$|$4) is merely  realized by embedding it in SU(2,2$|$4). 
 We would like to emphasize that the Killing vectors $R^{\Xi a}$ realize the Lie-algebra of PSU(2,2$|$4), given by (\ref{Liealgebra}), without the central charge. This is an advantage of the non-linear realization over the other two representations.

\subsection{PSU(2,2$|$4)/\{SO(1,4)$\otimes$SO(5)\}}

So far non-linear realization of PSU(2,2$|$4) has been discussed on the coset space PSU(2,2$|$ \ 4)/H without specifying a subgroup H\cite{Tsey}. Now we take H to be SO(1,4)$\otimes$ SO(6) to proceed with our discussions. First of all we note that 
$$
{\rm SU(2,2)}\otimes {\rm SU(4)}\cong{\rm SO(2,4)}\otimes {\rm SO(6)}
 \supset {\rm SO(1,4)}\otimes {\rm SO(5)}.
$$
The matrix representation of SU(2,2)$\otimes$SU(4) so far discussed can be identified with  the chiral spinor representation of SO(2,4)$\otimes$SO(6). The Dirac algebrae of SO(2,4) and SO(5) respectively  read 
\begin{eqnarray}
\{\Gamma_p,\Gamma_q\}&=&2\eta_{pq}=2(-1,\ 1\ ,\ 1\ ,\ 1\ ,\ 1\ ,-1 ),\quad\quad p,q=0,1,\cdots,5, 
\nonumber\\
\{\hat\Gamma_{\hat p},\hat\Gamma_{\hat q}\}&=& 2\hat\eta_{\hat p\hat q}=2(\ \hspace{0.1cm} 1\ ,\ 1\ ,\ 1\ ,\ 1\ ,\ 1\ ,\ 1\ ),\quad\quad \hat p, \hat q=0,1,\cdots,5. \nonumber
\end{eqnarray}
In the chiral spinor representation  the Dirac matrices  of SO(2,4) are, for example,  given by 
\begin{eqnarray}
\Gamma_5=\left(
\begin{array}{c|c}
 0 & 1 \\
\vspace{-0.4cm} & \\
\hline  
\vspace{-0.4cm} & \\
-1 & \hspace{0.15cm}0 \hspace{0.15cm}
\end{array}\right),\quad\quad
 \Gamma_m=\left(
\begin{array}{c|c}
 0 & \gamma_m \\
\vspace{-0.4cm} & \\
\hline  
\vspace{-0.4cm} & \\
 \gamma_m & 0 
\end{array}\right),     \label{gammamatrix}
\end{eqnarray}
with 4$\times$4 $\gamma$ matrices satisfying  
$$
\{\gamma_m,\gamma_n\}=2\eta_{mn}=2( -1\ ,\ 1\ ,\ 1\ ,\ 1\ ,\ 1\ ),\quad\quad m,n=0,1,2,\cdots,4. 
$$
On the other hand the Dirac matrices  of  SO(6) are given by 
\begin{eqnarray}
\hat\Gamma_5=\left(
\begin{array}{c|c}
\hspace{0.1cm} 0\hspace{0.2cm} &\hspace{0.2cm} 1\hspace{0.1cm} \\
\vspace{-0.4cm} & \\
\hline  
\vspace{-0.4cm} & \\
1 & 0 
\end{array}\right),\quad\quad\quad \hat\Gamma_{\hat m}=\left(
\begin{array}{c|c}
 0  & i\hat\gamma_{\hat m} \\
\vspace{-0.4cm} & \\
\hline  
\vspace{-0.4cm} & \\
\hspace{-0.15cm} -i\hat\gamma_{\hat m}\hspace{-0.15cm} & 0 
\end{array}\right), \label{gammamatrix'}
\end{eqnarray}
with 4$\times$4 $\hat\gamma$ matrices satisfying 
$$
\{\hat\gamma_{\hat m},\hat\gamma_{\hat n}\}=2\hat\eta_{\hat m\hat n}=2(\ 1\ ,\ 1\ ,\ 1\ ,\ 1\ ,\ 1\ ),\quad\quad \hat m,\hat n=1,2,\cdots,5.
$$
By using these Dirac matrices
the generators of SO(2,4) and SO(6) are given by
\begin{eqnarray}
\Gamma_{pq}={1\over 4}[\Gamma_p,\Gamma_q]{\rm P}_+, \quad\quad 
\hat\Gamma_{\hat p\hat q}={1\over 4}[\hat\Gamma_{\hat p},\hat\Gamma_{\hat q}]\hat {\rm P}_+. \nonumber
\end{eqnarray}
The chiral projectors take the diagonalized forms 
\begin{eqnarray}
{\rm P}_{+}={1\over 2}(1+\Gamma_7)=\left(
\begin{array}{c|c}
\hspace{0.1cm} 1\hspace{0.2cm} &\hspace{0.2cm} 0\hspace{0.1cm} \\
\vspace{-0.4cm} & \\
\hline  
\vspace{-0.4cm} & \\
0 & 0 
\end{array}\right),\quad\quad \hat {\rm P}_{+}={1\over 2}(1+\hat\Gamma_7)=
\left(
\begin{array}{c|c}
\hspace{0.1cm} 1\hspace{0.2cm} &\hspace{0.2cm} 0\hspace{0.1cm} \\
\vspace{-0.4cm} & \\
\hline  
\vspace{-0.4cm} & \\
0 & 0 
\end{array}\right),
 \nonumber
\end{eqnarray}
when  we choose  the Weyl representation 
\begin{eqnarray}
\gamma_0&=&\left(
\begin{array}{c|c}
\hspace{0.1cm} 0\hspace{0.2cm} &\hspace{0.2cm} 1\hspace{0.1cm} \\
\vspace{-0.4cm} & \\
\hline  
\vspace{-0.4cm} & \\
\hspace{-0.1cm}-1\hspace{-0.1cm} & 0 
\end{array}\right),\quad \gamma_1=\left(
\begin{array}{c|c}
\hspace{0.1cm} 0\hspace{0.2cm} &\hspace{0.2cm} \sigma_1\hspace{-0.1cm} \\
\vspace{-0.4cm} & \\
\hline  
\vspace{-0.4cm} & \\
\sigma_1 & 0 
\end{array}\right),\quad \gamma_2=\left(
\begin{array}{c|c}
\hspace{0.1cm} 0\hspace{0.2cm} &\hspace{0.2cm} \sigma_2\hspace{-0.1cm} \\
\vspace{-0.4cm} & \\
\hline  
\vspace{-0.4cm} & \\
\sigma_2 & 0 
\end{array}\right), \nonumber\\
\gamma_3&=&\left(
\begin{array}{c|c}
\hspace{0.1cm} 0\hspace{0.2cm} &\hspace{0.2cm} \sigma_3\hspace{-0.1cm} \\
\vspace{-0.4cm} & \\
\hline  
\vspace{-0.4cm} & \\
\sigma_3 & 0 
\end{array}\right), \quad \gamma_4=i\gamma_0\gamma_1\gamma_2\gamma_3,   \nonumber
\end{eqnarray}
for $\gamma_m$ in (\ref{gammamatrix}) and a similar representation  for $\gamma_{\hat m}$ in (\ref{gammamatrix'}).  
We have  one to one correspondence between the generators of SU(2,2)$\otimes$SU(4) in (\ref{set2}) and those of SO(2,4)$\otimes$O(6) as
\begin{eqnarray}
\{{\cal L}^\alpha_{\ \beta}\}&=&\{\Gamma_{12},\Gamma_{23},\Gamma_{31}\}, \quad\quad
\{P_{\alpha\dot\beta}\}=\{\Gamma_{\mu 4}+\Gamma_{\mu 5} \}, \nonumber\\
\{{\dot{\cal L}}^{\dot\alpha}_{\ \dot\beta}\}&=&\{\Gamma_{01},\Gamma_{02},\Gamma_{03}\},  
 \quad\quad  \{K_{\dot\alpha\beta}\}=\{\Gamma_{\mu 4}-\Gamma_{\mu 5} \}, \nonumber\\
&\ & \hspace{2.4cm}  {\cal D}=\Gamma_{45},    \nonumber\\
&\ & \hspace{1.8cm}\{R^a_{\ b}\}=\{\hat\Gamma_{\hat p\hat q}\}\nonumber
\end{eqnarray}
with $\mu=0,1,2,3$\cite{Aha}.

We further decompose the generators of SO(2,4)$\otimes$SO(6) under  SO(1,4)$\otimes$SO(5) as
\begin{eqnarray}
\Gamma_{pq}&=&\left[
\begin{array}{c|c}
 \Gamma_{mn} & \Gamma_{m5} \\
\vspace{-0.3cm} &  \\
\hline
\vspace{-0.3cm} &  \\
\Gamma_{5n}  & 0 \\
\end{array}\right], \quad\quad m,n=0,1,\cdots,4,     \nonumber\\
\hat \Gamma_{\hat p\hat q}&=&\left[
\begin{array}{c|c}
\hat  \Gamma_{\hat m\hat n} & \Gamma_{\hat m5} \\
\vspace{-0.3cm} &  \\
\hline
\vspace{-0.3cm} &  \\
\hat \Gamma_{5\hat n}  & 0 \\
\end{array}\right],  \quad\quad \hat m,\hat n=0,1,\cdots,4. \nonumber
\end{eqnarray}
Using this basis we rewrite the generators of PSU(2,2$|$4) in the matrix representation, given by (\ref{set2}),  as
\begin{eqnarray}
\{T^\Xi\}=\{\Gamma_{mn},\Gamma_{m5},\hat \Gamma_{\hat m\hat n},\hat \Gamma_{\hat m5}, {\cal S}^\alpha_{\ b},{\cal Q}^a_{\ \alpha},{\dot {\cal S}}^{a\dot\alpha},{\dot {\cal Q}}_{\dot\alpha a}\}.  \label{cosetgene}
\end{eqnarray}
Now we are in a position to construct the coset superspace PSU(2,2$|$4)/SO(1,4)$\otimes$SO(5), following the general method given previously.
The coset element $e^{i\phi\cdot T}$, given by (\ref{G/H}), takes an explicit form with   
\begin{eqnarray}
\{T^i\}&=&\{\Gamma_{m5},\hat \Gamma_{\hat m5},{\cal S}^\alpha_{\ b},{\cal Q}^a_{\ \alpha},{\dot {\cal S}}^{a\dot\alpha},{\dot {\cal Q}}_{\dot\alpha a}\}. \nonumber
\end{eqnarray}
These coset generators act on PSU(2,2$|$4)/SO(1,4)$\otimes$SO(5) transitively.  
They are identified with the corresponding generators of the $D=10$ Poincar\'e superalgebra  at the origin of the coset superspace. After this identification it is natural to rename the generators of $\{T^i\}$ as
\begin{eqnarray}
\{\Gamma_{m5},\hat \Gamma_{\hat m5}\} = \{\Gamma^M\},\quad\quad 
\{{\cal S}^\alpha_{\ b},{\cal Q}^a_{\ \alpha},{\dot {\cal S}}^{a\dot\alpha},{\dot {\cal Q}}_{\dot\alpha a}\}=\{{\cal Q}^{\frak M}\}, \label{renaming}
\end{eqnarray}
with $M=0,1,2,\cdots,9$ and ${\frak M}=1,2,\cdots,32$. 
Correspondingly  the coordinates  $\phi^{\bar i}$  repara\- metrizing PSU(2,2$|$4)/SO(1,4)$\otimes$SO(5)  are renamed  as
\begin{eqnarray}
\{\phi^{\bar i}\}=\{X^{\shortoverline M}, \Theta^{\shortoverline {\frak M}}\} \label{rename}.
\end{eqnarray}
They are identified with supercoordinates in the $D=10$ curved spacetime. 
By using them  we may write the coset element (\ref{G/H})  in a form looking like a vertex operator of the Green-Schwarz string theory as 
\begin{eqnarray}
e^{i\phi\cdot T}=e^{i(X\cdot \Gamma+\Theta\cdot {\cal Q})}.   \label{susycoordinates}
\end{eqnarray}
The Killing vectors defined by (\ref{Killing})  are  found as functions of  the the $D=10$ supercoordinates, 
\begin{eqnarray}
\delta^\Xi\phi^{\bar i}= R^{\Xi {\bar i}}(\phi)=\Big(R^{{\Xi\hspace{0.1cm}{\shortoverline M}}}(X,\Theta),R^{\Xi\hspace{0.1cm}\shortoverline{\frak M}}(X,\Theta)\Big).   \label{Killing2}
\end{eqnarray}

\section{Non-linear realization of Oscillators}
\setcounter{equation}{0}

In the previous section we have discussed that the coset element (\ref{susycoordinates}) looks like a vertex operator and  it transforms according to (\ref{nonlinear}), i.e.,  
\begin{eqnarray}
e^{i(X\cdot \Gamma+\Theta\cdot {\cal Q})}\longrightarrow e^{iX'(X,\Theta)\cdot \Gamma+\Theta'(X,\Theta)\cdot {\cal Q})} =e^{iM^\Xi T^\Xi}e^{i(X\cdot \Gamma+\Theta\cdot {\cal Q})}e^{-i\rho(X,\Theta,M)}, \label{transf3}
\end{eqnarray} 
in which  the non-linear transformations $X'(X,\Theta)$ and $\Theta'(X,\Theta)$ are  generated by the Killing vectors (\ref{Killing2}). The arguments have been given in  an algebraic way which does not relies on  either of the unitary(oscillator) representation and the matrix one. 
However let us now choose  the matrix representation. Then the transformation (\ref{transf3}) is written  by an $8\times 8$ supermatrix. 
If there also exists an 8-d column vector $\eta(X,\Theta)$ transforming as 
\begin{eqnarray}
\eta(X,\Theta) \longrightarrow \eta'(X',\Theta')=e^{i\rho(X,\Theta,M)}\eta(X,\Theta),  \label{eta}
\end{eqnarray}
 by the non-linear transformations $X'(X,\Theta)$ and $\Theta'(X,\Theta)$, 
then (\ref{transf3}) becomes 
\begin{eqnarray}
e^{i(X\cdot \Gamma+\Theta\cdot {\cal Q})}\eta(X,\Theta) \longrightarrow e^{iM^\Xi T^\Xi}e^{i(X\cdot \Gamma+\Theta\cdot {\cal Q})}\eta(X,\Theta).  \label{lineart}
\end{eqnarray}
It implies that $e^{i(X\cdot P+\Theta\cdot {\cal Q})}\eta(X,\Theta)$ is a covariant  vector under PSU(2,2$|$4). 
 In \cite{Ao1} such a quantity is called  Killing scalar $\Upsilon$, i.e., 
\begin{eqnarray}
\Upsilon(X,\Theta)= e^{i(X\cdot \Gamma+\Theta\cdot {\cal Q})}\eta(X,\Theta)).  \label{psi2}
\end{eqnarray}
The transformation  is exactly the same  as for the 8-d column vector 
\begin{eqnarray}
\psi=
\left(
\begin{array}{c}
a^\alpha \\
c^a  \\
b^\dagger_{\dot\alpha}
\end{array}
\right),    \label{psi4}  
\end{eqnarray}
which was defined by (\ref{psi}).  
Making the identification 
\begin{eqnarray}
 \Upsilon(X,\Theta) =\psi,  \label{psi3}
\end{eqnarray}
we claim that  this is a non-linear realization of the oscillators. 

The remaining question is whether  the quantity $\eta$ with the transformation property  (\ref{eta}) really exists.   In \cite{Ao1} the existence was shown  for the general bosonic coset space G/H in an arbitrary, but finite  representation of the  coset element $e^{i\phi\cdot T}$. We have chosen the matrix representation to discuss the coset space PSU(2,2$|$4)/\{SO(1,4)$\otimes$SO(5)\}. Therefore $\eta$  exists for this case similarly. 
Here we recall only of the point of the arguments and explain the quantity more explicitly for 
the coset space PSU(2,2$|$4)/\{SO(1,4)$\otimes$SO(5)\}.  First of all we consider  the Cartan-Maurer 1-form 
\begin{eqnarray}
g^{-1}dg=i(e^i_{\ \bar j}T^i+\omega^I_{\ \bar j}H^I)d \phi^{\bar j} \label{cartan},
\end{eqnarray}
denoting the coset element (\ref{susycoordinates}) as $g$ and using the index notation  (\ref{rename}). 
This defines the vielbein $e^i_{\ \bar j}$ and the connection $\omega^I_{\ \bar j}$ in the tangent frame of the coset space. Under the transformation (\ref{transf3}) they transform as 
\begin{eqnarray}
e^i_{\ \bar j}d\phi^{\bar j}&\longrightarrow &[e^{i\rho(\phi,M)}]^{ik}e^k_{\ \bar j}d\phi^{\bar j},  \nonumber\\
\omega^I_{\ \bar j}H^I d\phi^{\bar j}&\longrightarrow& e^{-i\rho(\phi,M)}[d -i\omega^I_{\ \bar j}H^Id\phi^{\bar j}]e^{i\rho(\phi,M)}. \nonumber
\end{eqnarray}
Then we have the Wilson line-operator 
$$
W(\phi,\phi_0)=P\exp i\int^\phi_{\phi_0}\omega^I_{\ \bar j}H^Id\phi^{\bar j},
$$
which transforms  as 
$$
W(\phi,\phi_0)\longrightarrow e^{i\rho(\phi,M)}W(\phi,\phi_0)e^{-i\rho(\phi_0,M)}.
$$
The compensator $e^{i\rho(\phi,M)}$ becomes a constant element at the origin $\phi=0$ of the coset space, i.e., 
\begin{eqnarray}
e^{i\rho(\phi,M)} \Big|_{\phi=0} =e^{iM^IH^I}\in {\rm H}={\rm SO(1,4)}\otimes{\rm SO(5)}.   \label{origin}
\end{eqnarray}
Let $e^{i M^I_0 H^I}\eta_0$ to be  a linear representation vector  with  $M^I_0$ parametrizing the subgroup H. Then it transforms by the compensator (\ref{origin}) at the origin as 
$$
 e^{iM^I_0H^I}\eta_0 \longrightarrow  e^{iM^IH^I}e^{M^I_0H^I}\eta_0.
$$ 
Here  $\eta_0$  is a constant vector  fixed in the representation space of H.
 To be concrete for the case of PSU(2,2$|$4)/\{SO(1,4)$\otimes$SO(5)\}, 
we have 
$$
e^{iM^I_0 H^I}\eta_0=e^{i(M^{mn}_0\Gamma_{mn}+M^{\hat m \hat n}_0{\hat \Gamma}_{\hat m \hat n})}\eta_0.
$$
by using the generators in (\ref{cosetgene}). 
Hence $\eta_0$  is now a constant chiral spinor of SO(1,4)$\otimes$SO(5). 
 As the result we find the quantity $\eta(X,\Theta)$
$$ 
\eta(\phi)=W(\phi,0)e^{iM_0^I H^I}\eta_0,
$$
with $\{\phi^{\bar i}\}=\{X^{\shortoverline M},\Theta^{\shortoverline {\frak M}}\}$, which has  the transformation property (\ref{eta}). 
   Thus we have justified  the identification  (\ref{psi2})  with  $\eta(X,\Theta)$ of this form.

\section{Exchange algebra}
\setcounter{equation}{0}

In the previous section we have identified the Killing scalar $\Upsilon(X,\Theta)$ of the coset space  PSU(2,2$|$4)/\{SO(1,4)$\otimes$SO(5)\} with 
 the 8-d column vector $\psi$ given by (\ref{psi4}). In \cite{Ao1} the general accounts for the Killing scalar were given for the ordinary coset space G/H, i.e., G is not a supergroup. It was shown that it  satisfies the classical exchange algebra of G in  the non-linear $\sigma$-model on G/H with the  Poisson brackets set up on the light-like line. For this it was essential to have the linear transformation property (\ref{lineart}), i.e.,  
\begin{eqnarray}
\delta^\Xi \Upsilon= T^\Xi \Upsilon, \label{linear}
\end{eqnarray}
by the Killing vectors (\ref{Killing2}). 
In this section we show that this is also true  for the Killing scalar (\ref{psi2}) of the non-linear $\sigma$-model on PSU(2,2$|$4)/\{SO(1,4)$\otimes$SO(5)\}.  The identification (\ref{psi3}) implies that the 8-d covariant vector $\psi$ given by (\ref{psi4}) satisfies  
the classical exchange algebra of PSU(2,2$|$4). Then the arguments go in the same way  for the most part even  for the coset superspace.  We shall here  explain them  taking a care of  the points for the supersymmetric generalization. 
 First of all we write the action of the non-linear $\sigma$-model on PSU(2,2$|$4)/\{SO(1,4)$\otimes$SO(5)\} 
\begin{eqnarray}
S= {1\over 2}\int d^2 x \ \eta^{+-}(e^i_{\ \bar j}\partial_+ \phi^{\bar j})(e^i_{\ \bar k}\partial_- \phi^{\bar k})(-1)^{g(i)},    \label{model}
\end{eqnarray}
with the vielbein $e^i_{\ \bar j}$ defined by (\ref{cartan}) and the supercoordinates $\phi^{\bar i}$ given by (\ref{rename}). Here we have the graded summation for the index $i$ according the quadratic Casimir (\ref{Casimir}). We set up  the Poisson brackets on the light-like line $x^+=y^+$ 
\begin{eqnarray}
&\ & \{\phi^{\bar i}(x)\mathop{,}^\otimes \phi^{\bar j}(y)\}=-{1\over 4 }\Big[\theta(x-y)t_{\Phi\Xi}^+ \delta^\Xi\phi^{\bar i}(x)\otimes \delta^\Phi\phi^{\bar j}(y) \nonumber\\
&\ &\hspace{3.7cm}-\theta(y-x)t_{\Phi\Xi}^+ \delta^\Xi\phi^{\bar j}(y)\otimes \delta^\Phi\phi^{\bar i}(x)(-1)^{g(\bar i)g(\bar j)}(-1)^{g(\Xi)g(\Phi)}\Big].\quad \label{Poisson}
\end{eqnarray}
The notation is as follows. $\theta(x)$ is the step function. $\delta^\Xi\phi^{\bar i}(x)$ are the Killing vectors defined by (\ref{Killing}). More correctly they should be written as $\delta^\Xi\phi^{\bar i}((\phi(x))$, but the dependence of $\phi^{\bar i}(x)$ was omitted to avoid an unnecessary complication. The quantity $t^+_{\Xi\Phi}$ is  the most crucial in our arguments. It is a modified Killing metric of $t_{\Xi\Phi}$. By means of it  we define  the classical r-matrix satisfying the classical Yang-Baxter equation. To explain this quantity  let us remember the definition of the classical r-matrices for the ordinary group 
\begin{eqnarray}
 r^{\pm}=\sum_{\alpha\in R} sgn\ \alpha E_{\alpha}\otimes E_{-\alpha} \pm
   \sum_{A,B} t_{AB}T^A \otimes T^B   
    \equiv  t^\pm_{\ AB}T^A\otimes T^B. \label{r}
\end{eqnarray}
Here  $T^A$ denote the generators of the group G with $t_{AB}$ the Killing metric. They are given in the Cartan-Weyl basis as $\{E_{\pm\alpha},H_\mu\}$ with $sgn\ \alpha=\pm$ according as the roots are positive or negative.
 Note the relation $t_{AB}^+=-t_{BA}^-$. The r-matrix satisfies the classical Yang-Baxter equation  
\begin{eqnarray}
 [r_{xy},r_{xz}]+[r_{xy},r_{yz}]+[r_{xz},r_{yz}]=0.   \nonumber
\end{eqnarray}
Here the r-matrix acts at  on a tensor product of the Killing scalars $\Upsilon(\phi(x))\otimes \Upsilon(\phi(y))\otimes \Upsilon(\phi(z))$ but only at the designated positions\cite{Ao1,Ao2}.   For the supergroup PSU(2,2$|$4) the r-matrix is generalized as follows. With the generators written as (\ref{set2}) we have  the quadratic Casimir (\ref{Casimir}), i.e., 
\begin{eqnarray}
 \sum_{\Xi\Phi}t_{\Phi\Xi}T^\Xi T^\Phi\equiv {\cal L}^\alpha_{\ \beta} {\cal L}^\beta_{\ \alpha}-{\cal R}^a_{\ b}{\cal R}^b_{\ a}+{\dot{\cal L}}^{\dot\alpha}_{\ \dot\beta} {\dot{\cal L}}^{\dot\beta}_{\ \dot\alpha}+{\cal D}^2 -\{T^{\alpha}_{\ \dot\beta},T^{\dot\beta}_{\ \alpha}\}-
[T^a_{\ \beta},T^\beta_{\ a} ]-[T^{a}_{\ \dot\beta},T^{\dot\beta}_{\ a}]. 
\nonumber
\end{eqnarray}
Correspondingly to this expression the r-matrix of PSU(2,2$|$4) is given by 
\begin{eqnarray}
 r^+&\equiv& \sum_{\Xi\Phi}t^+_{\Phi\Xi}T^\Xi T^\Phi \nonumber\\
&=& \sum_{\alpha>\beta}{\cal L}^\alpha_{\ \beta} {\cal L}^\beta_{\ \alpha}-\sum_{a>b}{\cal R}^a_{\ b}{\cal R}^b_{\ a}+\sum_{\dot\alpha>\dot\beta} {\dot{\cal L}}^{\dot\alpha}_{\ \dot\beta} {\dot{\cal L}}^{\dot\beta}_{\ \dot\alpha}-\sum_{{\rm all}\ \alpha,\dot\beta}T^{\alpha}_{\ \dot\beta}T^{\dot\beta}_{\ \alpha}-\sum_{{\rm all}\ a\beta}T^a_{\ \beta}T^\beta_{\ a}-\sum_{{\rm all}\ a\dot\beta}T^{a}_{\ \dot\beta}T^{\dot\beta}_{\ a} \nonumber\\
&\ &  + \sum_{\Xi\Phi}t_{\Phi\Xi}T^\Xi T^\Phi.   \nonumber
\end{eqnarray}
That is,  $t^+_{\Xi\Phi}$ in (\ref{Poisson}) is a simple generalization of the quantity in (\ref{r})  for the case of  
PSU(2,2$|$4). It is straightforward to show that the  r-matrix generalized in this way satisfies the classical Yang-Baxter equation
\begin{eqnarray}
 [r_{xy},r_{xz}\}+[r_{xy},r_{yz}\}+[r_{xz},r_{yz}\}=0,   \label{gCYB}
\end{eqnarray}
with the graded commutator [\ ,\ \}. 
Note that now we have 
$$
t_{\Xi\Phi}=(-1)^{g(\Xi)g(\Phi)}t_{\Phi\Xi},\quad\quad t^+_{\Xi\Phi}=-(-1)^{g(\Xi)g(\Phi)}t^-_{\Phi\Xi}.
$$
Then it follows that  
$$
\{\phi^{\bar i}(x)\mathop{,}^\otimes \phi^{\bar j}(y)\}
=(-1)^{g(\bar i)g(\bar j)}\{\phi^{\bar j}(y)\mathop{,}^\otimes \phi^{\bar i}(x)\},
$$
for the Poisson brackets given by (\ref{Poisson}). All the arguments here  on  the classical Yang-Baxter equation were 
 done  for the supergroup SL(1$|$2) and OSP(2$|$2) in \cite{Ao4,Ao5}. There the r-matrix of  the respective supergroup appeared   as showing integrability of the $D=2$, $(1,0)$ and (2,0) effective gravity. 

Finally we can show the consistency of the Poisson brackets (\ref{Poisson}). First of all it satisfies the Jacobi identities owing to the classical Yang-Baxter equation for the r-matrix. Secondly the energy-momentum tensor of the non-linear $\sigma$-model (\ref{model}) reproduces the diffeomorphism  
\begin{eqnarray}
\delta_{diff} \phi^{\bar i}(x^+,x^-) &\equiv&\epsilon(x^-)\partial_- \phi^{\bar i}(x^+,x^-)  \nonumber\\
          &=& \int dy^- \epsilon(y^-)\{\phi^{\bar i}(x), T_{--}(\phi(y))\}\Big |_{x^+=y^+},
 \nonumber
\end{eqnarray}
Thirdly the Poisson brackets tend to those of the free boson and fermion theory as 
$$
\{\phi^{\bar i}(x)\mathop{,}^\otimes \phi^{\bar j}(y)\}
=-{1\over 4}[\theta(x-y)\delta^{\bar i\bar j} - \theta(y-x) \delta^{\bar j\bar i} (-1)^{g(\bar i)g(\bar j)}].
$$
These statements can be verified in the same way as for the ordinary non-linear $\sigma$-model. 

With the Poisson brackets (\ref{Poisson}) let us  calculate 
$\displaystyle{\{\Upsilon(x)\mathop{,}^\otimes \Upsilon(y)\}}$ for the Killing scalar  $\Upsilon$ using the property 
\begin{eqnarray}
\{\phi^{\bar i}(x)\mathop{,}^\otimes \Upsilon(y)\}=\{\phi^{\bar i}(x)\mathop{,}^\otimes \phi^{\bar j}(y)\}{\delta \Upsilon(y)\over \delta \phi^{\bar j}(y)},  \nonumber
\end{eqnarray}
together with (\ref{linear}). We then get  the classical exchange algebra in the form
\begin{eqnarray}
\{\Upsilon(x)\mathop{,}^\otimes \Upsilon(y)\}=-{1\over 4}[\theta(x-y)r^+ + \theta(y-x) r^-]\Upsilon(x)\otimes\Upsilon(y),  \label{cea} 
\end{eqnarray}
on the light-like plane $x^+=y^+$. 
Here $\Upsilon(x)$ should be understood with an abbreviated notation for $\Upsilon(\phi(x))$. It is a non-linear realization of  the oscillators by the identification (\ref{psi2}). Thus the oscillators obey the classical exchange algebra (\ref{cea}).  

The supergroup PSU(2,2$|$4) is a simple group. Hence we may use  the plug-in formula to promote the r-matrix  to the universal R-matrix $R_{xy}$. It is  expressed purely in terms of generators of G. Then (\ref{cea}) becomes 
\begin{eqnarray}
\Upsilon(x)\otimes \Upsilon(y)= \theta(x-y)R_{xy}^+ \Upsilon(y)\otimes \Upsilon(x) +
\theta(y-x)R_{xy}^- \Upsilon(y)\otimes \Upsilon(x).    \label{QEA2}
\end{eqnarray} 
Here the universal R-matrix satisfies  the quantum Yang-Baxter equation
\begin{eqnarray}
R_{xy}R_{xz}R_{yz}=R_{yz}R_{xz}R_{xy}.  \label{QQQ}
\end{eqnarray}
 (\ref{cea}) and (\ref{gCYB}) are the respective  classical correspondents
 of (\ref{QEA2}) and (\ref{QQQ})  obtained   by 
$$
R_{xy}^\pm=1 +hr_{xy}^\pm+O(h^2),  
$$
with $h=-{1\over 4}$. 
 For G=SU(2) or SU(1,1) the plug-in formula of the universal R-matrix  is given in a rather simple form as 
\begin{eqnarray}
R=q^{{1\over 2}H\otimes H}\sum_{n=0}^\infty {q^{{1\over 2}n(n-1)}(q-q^{-1})^n\over [n]_q!}({ E}^+)^n\otimes ({E}^-)^n
= \exp_q[(q-q^{-1}){E}^+\otimes { E}^-],   \label{R}
\end{eqnarray}
by using the notation (\ref{r}) for the Lie-algebra. Here the $q$-exponential 
is defined by 
$$
\exp_q(x)=\sum_{n=0}^\infty {q^{{1\over 2}n(n-1)}x^n\over [n]_q!}, 
$$
with $q=e^h$ and 
$$
[n]_q={q^n-q^{-n}\over q-q^{-1}}.
$$
The formula (\ref{R}) can be generalized for the general simple group. 
The generalized plug-in formula for the ordinary group can be found in \cite{Pres}. That for supergroups 
 was given in \cite{Kho}. 

We consider a correlation function 
\begin{eqnarray}
<{\cal V}_1(x_1)\otimes {\cal V}_2(x_2)\otimes\cdots\otimes {\cal V}_i(x_i)\otimes{\cal V}_{i+1}(x_{i+1})\otimes\cdots \otimes{\cal V}_N(x_N)>.  \label{correlation}
\end{eqnarray}
Here ${\cal V}_i(x_i)$  are the ${\cal N}=4$ SUSY field strength  multiplet and  their descendants  in table 1, i.e.,  
$$
\{{\cal V}(x)   \}=\{{\frak D}^kF(x),{\frak D}^k\Psi(x),{\frak D}^k\Phi(x),{\frak D}^k\dot\Psi(x), {\frak D}^k\dot F(x)\}, \quad k=0,1,2,\cdots.
$$
They are arrayed on the light-like line $x^+_1=x^+_2=\cdots =x^+_N$ as shown in figure 2.  

\vspace{0.5cm}

\begin{figure}[bh]
  \begin{center}
   \includegraphics[angle=0, width=0.5\textwidth]{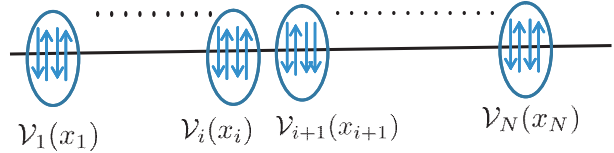}
  \end{center}
\vspace{-0.2cm}
\caption{Arrows stand for SU(2)-spins of the oscillators $a^\alpha,b^{\dot\alpha},c^{\bar a},d^{\dot a}$. Encircling them indicates a component of the ${\cal N}=4$ field strength multiplet ${\cal V}_i(x_i)$.}

 \end{figure}

\noindent
 Each of the multiplets is given in terms of the oscillators  as in table 2. Let ${\cal V}_i(x_i)$ and ${\cal V}_{i+1}(x_{i+1})$ at the adjacent positions to be one of the components belonging to ${\frak D}_{\alpha\dot\beta}\Phi$ and $\Psi$ respectively. For instance,  we  have  
\begin{eqnarray}
{\cal V}_i(x_i)&=& (a^\dagger_\alpha(x_i) b^\dagger_{\dot\alpha}(x_i) )(c^\dagger_{\bar a} (x_i) d^\dagger_{\dot a}(x_i)), \nonumber\\  {\cal V}_{i+1}(x_{i+1})&=& a^\dagger_\alpha(x_{i+1}) c^\dagger_{\bar a}(x_{i+1}) d^\dagger_3(x_{i+1})d^\dagger_4(x_{i+1}),    \nonumber
\end{eqnarray}
with ${\frak D}_{\alpha\dot\beta}=P_{\alpha\dot\beta}=a^\dagger_\alpha b^\dagger_{\dot\beta}$. The quantum exchange algebra  for the field strength multiplet ${\cal V}_i(x_i)$ follows  from  (\ref{QEA2}), because the oscillators are identified with   the Killing scalar $\Upsilon$ as (\ref{psi3}). 
  That is, the quantum exchange algebra  is obtained by braiding the respective oscillators in ${\cal V}_i(x_i)$ and ${\cal V}_j(x_j)$ one by one. The R-matrix is now in an infinite-dimensional representation.

The ${\cal N}=4$ SUSY field strength multiplet ${\cal V}$ is the simplest one. 
The  arguments  can be similarly applied  to other  multiplets than ${\cal N}=4$ SUSY field strength multiplet ${\cal V}$. A summary of the general  superconformal multiplets  is given in appendix A.

\section{Conclusions}
\setcounter{equation}{0}

Thus we are led to conclude that the supercoherent space of the unitary(oscillator) representation of PSU(2,2$|$4) becomes non-commutative. To show this, we have constructed the oscillators  as the Killing scalar $\Upsilon$ in the non-linear $\sigma$-model on  PSU(2,2$|$4)/\{SO(1,4)$\otimes$SO(5)\}. It was argued  that they  satisfy the exchange algebra when the model  is quantized on the light-like line. 
 They  took  a suggestive form of the vertex operator of the  Green-Schwarz superstring  as given by (\ref{psi2}), i.e., 
\begin{eqnarray}
\Upsilon=e^{i(X\cdot P+\Theta\cdot {\cal Q})}\eta(X,\Theta)).    \nonumber
\end{eqnarray}

We comment on the $D=10$ flat space-time limit of the non-linear $\sigma$-model
on PSU(2,2$|$4)/\{SO(1,4)$\otimes$SO(5)\}. Taking a naive limit where the AdS$_5\otimes$S$^5$ radius tends to $\infty$ does not give the desired $D=10$ super-Poincar\'e invariance to the non-linear $\sigma$-model. According to \cite{Ber} a correct way to go to the the flat space-time limit is to rescale the structure constants  in  the algebrae  (\ref{L+-} ) as 
\begin{eqnarray}
\{S^\alpha_{\ b},Q^a_{\ \beta}\}&=& {1\over r}\delta^a_b L^\alpha_{\ \beta}+{1\over r}\delta^\alpha_{\beta}R^a_{\ b} + {1\over 2}\delta^a_b\delta^\alpha_{\beta}(D-C),   
 \nonumber\\
\{{\dot S}^{a\dot\beta}, {\dot Q}_{\dot\alpha b}\}&=& {1\over r}\delta^a_b{\dot L}^{\dot\beta}_{\ \dot\alpha} - {1\over r}\delta_{\dot\alpha}^{\dot\beta}R^a_{\ b}+{1\over 2}\delta^a_b\delta_{\dot\alpha}^{\dot\beta}(D+C).   \nonumber 
\end{eqnarray}
Then in the limit $r\rightarrow \infty$ the vielbein defined by (\ref{cartan}) tends to 
\begin{eqnarray}
e^M_{\ \ \bar j}d\phi^{\overline j}=d X^{M}-\Theta\cdot \Gamma^M d \Theta, \quad\quad\quad
 e^{\frak M}_{\ \bar j}d\phi^{\bar j}= d \Theta^{{\frak M}},  \label{vielbein}
\end{eqnarray} 
 by using (\ref{renaming}) and (\ref{rename}). We no longer need  distinguish the coordinates of the coset space and those of the tangent space, so that  
$$
 X^{\shortoverline M}=X^M,\quad\quad \Theta^{\shortoverline{\frak M}}=\Theta^{\frak M}.
$$
The vielbeins in  (\ref{vielbein})  are invariant under the global supertransformation
$$
\delta X^M=\alpha\cdot \Gamma^M\Theta, \quad\quad \delta\Theta^{\frak M}=\alpha^{\frak M}.
$$

We also comment on the fact that our correlation function (\ref{correlation}) is independent of the position $x_i$ of the observables ${\cal V}_i(x_i)$ on the light-like line. It might be considered as a correlation function  of the similar kind to the one   in the topological field theory. The exchange algebra of OSP(2$|$2) for the (2,0) topological gravity was discussed in such a context in \cite{Ao5}.  
We may think of position-dependence for  the correlation function such as 
\begin{eqnarray}
e^{ip_1x_1}e^{ip_2x_2}\cdots e^{ip_Nx_N}<{\cal V}_1(x_1)\otimes {\cal V}_2(x_2)\otimes\cdots\cdots \otimes{\cal V}_N(x_N)>, \label{cor}
\end{eqnarray}
in which  $p_1,p_2,\cdots,p_N$ are momenta excited along the light-like line. But PSU(2,2$|$4) is too restrictive to allow for  such a dependence.
 Therefore we think of breaking the symmetry of PSU(2,2$|$4) to a subgroup symmetry which contains a certain number of central charges as
 factor groups.  In \cite{Bei} they took such a subgroup to be PSU(1,1$|$2)$\ltimes\mathbb{R}^3$.  The generators $D,P,K$ in (\ref{matrix}) are reduced to 
 three U(1) charges acting on the correlation function (\ref{cor}). 
The position-dependent R-matrix acting  on the correlation function (\ref{cor})  was given for this residual subgroup in \cite{Bei}. 
It played a crucial role in discussing  the duality between the $D=2$ spin-chain  and the $D=4, \ {\cal N}=4$ SUSY YM theory.
It  is interesting  to investigate how such a position-dependent R-matrix occurs as symmetry  breaking of the universal R-matrix of PSU(2,2$|$4) in the non-linear $\sigma$-model. The issue will be discussed  in a forthcoming publication\cite{Ho}.

\vspace{2cm}

\appendix
\section{${1\over 4}$- and ${1\over 8}$-BPS multiplets } 
\setcounter{equation}{0}

In section 3 we have argued that  the ${\cal N}=4$ SUSY field strength multiplet is the smallest BPS multiplet 
 of the unitary(oscillator) representation of PSU(2,2$|$4). In this appendix  we continue the arguments\cite{Bei3,Doh,Ferr}  to give larger BPS multiplets representing  other ${\cal N}=4$ superconformal multiplets. Our arguments on the exchange algebra 
  in this paper can be straightforwardly applied to those multiplets as well. 

In order to enlarge the BPS multiplet discussed in section 3, we retain  only half of the supercharges in (\ref{1/2}) 
\begin{eqnarray}
Q^1_{\ \beta}, \quad\quad {\dot Q}_{\dot\alpha 4},     \nonumber
\end{eqnarray}
as  fermionic generators annihilating the vacuum. 
They break the R-symmetry SU(2)$\otimes$ SU(2) of the physical vacuum of the ${1\over 2}$-BPS multiplet    to U(1)$\otimes$ U(1). 
To be explicit, they are 
\begin{eqnarray}
a^\dagger_\beta c^1, \quad\quad b^\dagger_{\dot\beta}d^4.
   \nonumber
\end{eqnarray}
Then the  following states 
\begin{eqnarray}
 Z,   \quad\quad Y=c^\dagger_2 d^\dagger_3Z,      \nonumber
\end{eqnarray}
 are annihilated by them  and satisfies the constraint $C=0$. This defines  the physical vacuum of the ${1\over 4}$-BPS multiplet. 
 If we further halve the number of the annihilating supercharges  and retain only 
\begin{eqnarray}
 {\dot Q}_{\dot\alpha 4}    
 \nonumber
\end{eqnarray}
the R-symmetry gets enlarged as  SU(3)$\otimes$ U(1).  The states  which are annihilated by them  and satisfy $C=0$ are 
\begin{eqnarray}
 Z, \quad\quad Y=c^\dagger_2 d^\dagger_3Z,\quad\quad X=c^\dagger_1 d^\dagger_3Z.   \label{ZYX}
\end{eqnarray}
This defines  the vacuum of  the  ${1\over 8}$-BPS multiplet.

Let us denote the fermionic oscillators belonging  to the representation ${\bf 4}^*$ of the R-symmetry SU(4) by the Dynkin label as
\begin{eqnarray}
c^\dagger_1= [-1,0,0], \quad\quad   
c^\dagger_2= [1,-1,0], \quad\quad  
d^3= [0,1,-1],  \quad\quad 
d^4= [0,0,1].  \nonumber
\end{eqnarray}
 Then the states $X,Y,Z$ in (\ref{ZYX}) are denoted by the Dynkin label as
\begin{eqnarray}
Z= [0,1,0], \quad\quad   
Y= [1,-1,1], \quad\quad  
X= [-1,0,1],  \quad\quad   \nonumber
\end{eqnarray}
which form the representation ${\bf 3}^*$ of the R-symmetry SU(3). 
The states of the representation ${\bf 3}$  are 
\begin{eqnarray}
Z^*= [0,-1,0], \quad\quad   
Y^*= [-1,1,-1], \quad\quad  
X^*= [1,0,-1].  \quad\quad   \nonumber
\end{eqnarray}
In terms of the oscillators they are given by 
$$
Z^*=c^\dagger_1c^\dagger_2d^\dagger_3d^\dagger_4Z, \quad\quad  
Y^*=c^\dagger_1d^\dagger_4Z, \quad\quad 
X^*= c^\dagger_2d^\dagger_4Z.
$$
Note that these six states were given by the field $\Phi$ in table 2.


\begin{table}[t]
\begin{center}
\begin{tabular}{|c|c|cc|cc|cc|} 
 \hline
\lw {type} & \lw {$(\sharp Q,\sharp \dot Q)$} & \multicolumn{2}{c|}{SU(2)$\otimes$SU(2)} & \multicolumn{2}{c|} {SU(2)$\otimes$SU(2)$\otimes$SU(4)} & \multicolumn{2}{c|}{h.w.}     \\
\vspace{-0.6cm}                                                             \\
     &         &\multicolumn{2}{c|}{h.w.}& \multicolumn{2}{c|}{h.w.}    & \multicolumn{2}{c|}{dimension $\Delta$}    \\ 
\hline 
${1\over 2}$-BPS  & (4,4)  & \hspace{0.7cm} $[j+2,\bar j +2]$  &&\hspace{0.3cm} $[j,\bar j][0,k,0],\ k>0$    &&  $k$  & \\ 
${1\over 4}$-BPS  & (2,2)  & \hspace{0.7cm}  $[ j+3,\bar j+3]$  &&\hspace{0.3cm} $[j,\bar j][l,k,l],\ l>0$    &&  $k+2l$   &   \\
${1\over 8}$-BPS  & (2,0)  & \hspace{0.7cm}  $[3,\bar j+4]$  &&\hspace{0.3cm} $[0,\bar j][l,k,l+2m],\ m>0$ &&  $k+2l+3m$  & \\  
\hline     
\end{tabular}
\end{center}
\vspace{-0.2cm}
\caption{The BPS multiplets. $\sharp Q$ and $\sharp \dot Q$  are the numbers of supercharges annihilating the highest weight state. The third column indicates the highest weight state of SU(2)$\otimes$SU(2), excited from the highest state in the fourth column by applying the remaining supercharges.   Further excitation is possible by applying the space-time derivative $P_{\alpha\dot\beta}$. $\Delta$ in the last column  is the conformal  dimension  defined   by the dilatation  in (\ref{U(1)}). }

\end{table}

By a systematic analysis of the unitary representation of PSU(2,2$|$4) \cite{Bei3,Doh,Ferr} the BPS multiplets are known as given in table 3\cite{Doh}. The physical vacua of the BPS multiplets are 
 the highest weight states of the unitary representation of PSU(2,2$|$4). They are  given  by  the following tensor  products 
\begin{eqnarray}
Z^k\quad\quad &{\rm for }&\quad {1\over 2}\ {\rm BPS}, \hspace{2cm}  \nonumber \\
 Y^lZ^{k+l}\quad\quad &{\rm for }&\quad {1\over 4}\ {\rm BPS},   \label{BPS} \\
X^mY^{l+m}Z^{l+m+k}\quad\quad &{\rm for }&\quad {1\over 8}\ {\rm BPS},   \nonumber 
\end{eqnarray}
with (anti-)symmetrization indicated by  the SO(6) Young tableau in figure 3. 
Here we have to take $k$ copies of the  oscillators $a^\alpha,b^{\dot \alpha},c^a$ and $a^\dagger_\alpha,b^\dagger_{\dot \alpha},c^\dagger_a$. Correspondingly the Fock space which they act on 
 is  the tensor product of $k$ copies of the Fock space (\ref{Fock2}).  For instance, we understand 
 $Z$, given by (\ref{ZZ}), as $\sum_{s=1}^k c^{(s)\dagger}_3c^{(s)\dagger}_4$. It is easy to verify that the physical vacua in (\ref{BPS}) are indeed  the SU(4) highest weight states  and  the conformal dimension $\Delta$, respectively given  in table 3.

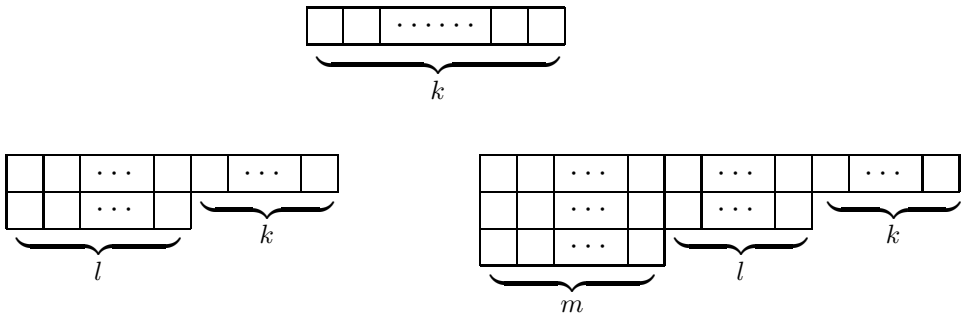
\begin{figure}[b]

\vspace{1cm}

\setlength{\unitlength}{0.7mm}
\begin{picture}(70,28)(-20,0)

\put(57,35){\line(0,-1){7}}
\put(64,35){\line(0,-1){7}}
\put(71,35){\line(0,-1){7}}
\put(92,35){\line(0,-1){7}}
\put(99,35){\line(0,-1){7}}
\put(106,35){\line(0,-1){7}}

\put(57,35){\line(1,0){49}}
\put(57,28){\line(1,0){49}}

\put(75,28){\makebox(7,7){\quad$\cdots\cdots$}}

\put(57,26){
$\underbrace{\makebox(46,10){}}_{\mbox{\footnotesize{$k$}}}$
}

\put(63,7){\line(0,-1){7}}
\put(56,7){\line(0,-1){7}}
\put(42,7){\line(0,-1){7}}
\put(35,7){\line(0,-1){7}}
\put(35,7){\line(0,-1){14}}
\put(28,7){\line(0,-1){14}}
\put(14,7){\line(0,-1){14}}
\put(7,7){\line(0,-1){14}}
\put(0,7){\line(0,-1){14}}

\put(0,7){\line(1,0){63}}
\put(0,7){\line(1,0){63}}
\put(0,0){\line(1,0){63}}
\put(0,-7){\line(1,0){35}}

\put(14,0){\makebox(7,7){\quad$\cdots$}}
\put(42,0){\makebox(7,7){\quad$\cdots$}}
\put(14,-7){\makebox(7,7){\quad$\cdots$}}

\put(0,-8.5){
$\underbrace{\makebox(31,10){}  }_{\mbox{\footnotesize{$l$}}}$
}

\put(35,-1.5){
$\underbrace{\makebox(25,10){}  }_{\mbox{\footnotesize{$k$}}}$
}

\put(181,7){\line(0,-1){7}}
\put(174,7){\line(0,-1){7}}
\put(160,7){\line(0,-1){7}}
\put(153,7){\line(0,-1){14}}
\put(146,7){\line(0,-1){14}}
\put(132,7){\line(0,-1){14}}
\put(125,7){\line(0,-1){14}}
\put(125,7){\line(0,-1){21}}
\put(118,7){\line(0,-1){21}}
\put(104,7){\line(0,-1){21}}
\put(97,7){\line(0,-1){21}}
\put(90,7){\line(0,-1){21}}

\put(90,7){\line(1,0){91}}
\put(90,0){\line(1,0){91}}
\put(90,0){\line(1,0){63}}
\put(90,-7){\line(1,0){63}}
\put(90,-14){\line(1,0){35}}

\put(104,0){\makebox(7,7){\quad $\cdots$}}
\put(132,0){\makebox(7,7){\quad$\cdots$}}
\put(160,0){\makebox(7,7){\quad$\cdots$}}
\put(104,-7){\makebox(7,7){\quad$\cdots$}}
\put(132,-7){\makebox(7,7){\quad$\cdots$}}
\put(104,-14){\makebox(7,7){\quad$\cdots$}}

\put(90,-15.5){
$\underbrace{\makebox(31,3){}  }_{\mbox{\footnotesize{$m$}}}$
}

\put(125,-8.5){
$\underbrace{\makebox(25,3){}  }_{\mbox{\footnotesize{$l$}}}$
}

\put(154,-1.5){
$\underbrace{\makebox(25,3){}  }_{\mbox{\footnotesize{$k$}}}$
}

\end{picture}

\vspace{2cm}

\caption{The SO(6) Young tableaux representing symmetrization and anti-symmetrization for the ${1\over 2}$-,${1\over 4}$-,${1\over 8}$-BPS states.  }

\end{figure}

The BPS multiplets with $\Delta\ge 2$, given in table 3, are  other  ${\cal N}=4$ SUSY superconformal  multiplets than the field strength multiplet in table 1. 
So far our arguments have been done by assuming the gauge group to be Abelian. 
 All the fields  of ${\cal N}=4$ superconformal  multiplet are in the adjoint representation of a gauge group. Let them to be Lie-valued in the gauge 
  algebra. Then the physical vacua in (\ref{BPS}) are given as the gauge singlets 
\begin{eqnarray}
tr[Z^k]\quad\quad &{\rm for }&\quad {1\over 2}{\rm\mathchar`-BPS}, \hspace{3.5cm}  \nonumber \\
tr[Y^l]tr[Z^{k+l}]\quad\quad &{\rm for }&\quad {1\over 4}{\rm \mathchar`-BPS},   \label{physicalv} \\
tr[X^m]tr[Y^{l+m}]tr[Z^{l+m+k}]\quad\quad &{\rm for }&\quad {1\over 8}{\rm \mathchar`-BPS}.   \nonumber 
\end{eqnarray}
Here $tr$ denotes trace over the gauge algebra. Therefore we have $k,l,m \ge 2$ for a simple gauge group. We may consider other multitrace states obtained  by replacements such as 
\begin{eqnarray}
tr[Z^k]\longrightarrow  tr[Z^{k-l}]tr[Z^l], \quad\quad
tr[Y^k]tr[Z^l]\longrightarrow tr[Y^kZ^l], \quad\quad etc  
 \label{replace}
\end{eqnarray}
since they  do not change the dimension $\Delta$. However we have 
$$
tr[Z^k]= tr[Z^{k-l}]tr[Z^l],
$$
by the symmetrization indicated by the SO(6) Young tableau in figure 3. 
 Therefore the former replacement gives nothing new[].  On the other hand the anti-symmetrization by the  SO(6) Young tableau  implies that  
$$
tr[Y^l]tr[Z^k]\ne tr[Y^lZ^k].
$$
That is, the latter replacement of (\ref{replace}) changes the physical vacuum into a non-BPS vacuum,  Thus  the multitrace states with multiplicity greater than $3$ are irrelevant for the BPS multiplets in table 3.

Finally it is interesting to understand  the vacua of the BPS multiplets so far discussed in a  way which manifests the SO(6) symmetry of the Young tableau\cite{Ferr}. 
We consider the generators  $Q^a_{\ \alpha}$ and $\dot Q_{\dot\alpha a}$ and make  three sets as
$$
\{ {Q}^1_{\ \alpha},  {Q}^2_{\ \alpha}, \dot{Q}_{\dot\alpha 3}, \dot{Q}_{\dot\alpha 4}\},\quad\quad
\{ {Q}^1_{\ \alpha},  {Q}^3_{\ \alpha}, \dot{Q}_{\dot\alpha2}, \dot{Q}_{\dot\alpha 4}\},
\quad\quad
\{ {Q}^2_{\ \alpha},  {Q}^3_{\ \alpha}, \dot{Q}_{\dot\alpha 1}, \dot{Q}_{\dot\alpha 4}\},
$$
in each of which all the generators are anti-commuting. Then the states  $X,Y,Z$ in (\ref{ZYX}) satisfy constraints  as 
\begin{eqnarray}
 {Q}^1_{\ \alpha}{Z}&=& {Q}^2_{\ \alpha} {Z} =\dot{Q}_{\dot\alpha 3}{Z}=\dot{Q}_{\dot\alpha 4}{Z}=0, \nonumber\\
 {Q}^1_{\ \alpha} {Y}&=&{Q}^3_{\ \alpha} {Y}=\dot{Q}_{\dot\alpha 2}{Y}=
\dot{Q}_{\dot\alpha 4}{ Y}=0,  \nonumber\\
{Q}^2_{\ \alpha} {X}&=&{Q}^3_{\ \alpha} {X}=\dot{Q}_{\dot\alpha 1}{X}=
\dot{Q}_{\dot\alpha 4}{ X}=0.  \nonumber
\end{eqnarray}
Consequently  the physical vacua (\ref{physicalv}) to be constrained by
\begin{eqnarray}
{Q}^1_{\ \alpha}tr[{Z}^k]= {Q}^2_{\ \alpha} tr[{Z}^k]=\dot{Q}_{\dot\alpha 3}tr[{Z}^k]=\dot{Q}_{\dot\alpha 4}tr[{Z }^k]=0 
\quad\quad &{\rm for }&\quad{1\over 2}{\rm \mathchar`-BPS},  \nonumber \\
{Q}^1_{\ \alpha}\Bigg(tr[{Y}^l]tr[{Z}^{k+l}]\Bigg)=\dot{Q}_{\dot\alpha 4}\Bigg(tr[{Y}^l]tr[{Z}^{k+l}]\Bigg)=0
\quad\quad &{\rm for }&\quad {1\over 4}{\rm \mathchar`-BPS},   \nonumber \\
\dot{Q}_{\dot\alpha 4}\Bigg(tr[{X}^m]tr[{Y}^{l+m}]tr[{Z}^{l+m+k}]\Bigg)=0 \quad\quad &{\rm for }&\quad {1\over 8}{\rm \mathchar`-BPS}.   \nonumber 
\end{eqnarray}
They are nothing but the constraints we have so far discussed.

\section{Algebraic calculation of the Killing vectors} 
\setcounter{equation}{0}

The Killing vectors of the coset space G/H  are defined when G is simple or most generally speaking  semi-simple. 
 The supergroup PSU(2,2$|$4) is simple. But  either of the unitary(oscillator)  and the matrix representation discussed in this paper  realizes it  as a  subgroup of SU(2,2$|$4) which is not simple. In this appendix we present a purely algebraic way  to calculate the Killing vectors G/H, which is free from the central charge of SU(2,2$|$4).

When the parameters $M$ are infinitesimally small,  we may write the  transformation (\ref{nonlinear}) as
\begin{eqnarray}
 e^{i[\phi\cdot T+M^\Xi R^\Xi\cdot T + O(M^2)]}
= e^{iM^\Xi T^\Xi}e^{i\phi\cdot T}e^{-i[M^\Xi\rho^\Xi\cdot H]},  \label{transf}
\end{eqnarray}
Here use was made of the definition of the Killing vectors (\ref{Killing}) and  the notation 
\begin{eqnarray}
R^\Xi(\phi)\cdot T  \equiv  R^{\Xi \bar i}(\phi)T^{i}, \quad\quad 
\rho^\Xi(\phi)\cdot T \equiv   \rho^{\Xi\bar I}(\phi,M)H^I. 
\nonumber 
\end{eqnarray}
(\ref{transf}) becomes
\begin{eqnarray}
&\ & e^{i[\phi\cdot T+M^\Xi R^\Xi\cdot T + O(M^2) ]}  \nonumber\\
&\ &\hspace{1cm}= e^{i[\phi\cdot T + \sum_{n=0}^\infty \alpha_n(ad\ i\phi\cdot T)^n (M^\Xi T^\Xi) - \sum_{n=0}^\infty(-1)^n \alpha_n(ad\ i\phi\cdot T)^n 
(M^\Xi\rho^\Xi\cdot H )
+ O(\epsilon^2))]},  \label{non-def1}
\end{eqnarray}
by using the following formulae: for  matrices  ${\cal E}$ and $X$
\begin{eqnarray}
\exp{\cal E}\exp X &=& \exp \biggl( X+\sum_{n=0}^\infty \alpha_n(ad\ X)^n{\cal E} +   O({\cal E}^2)   \biggr),    \label{formula1}  \\
\exp X\exp{\cal E} &=& \exp \biggl( X+\sum_{n=0}^\infty (-1)^n\alpha_n(ad\ X)^n{\cal E} +  
 O({\cal E}^2)   \biggr), \label{formula2}
\end{eqnarray}
if ${\cal E} \ll 1$. Here $\alpha_n$ are the constants recursively determined by 
\begin{eqnarray}
\alpha_n +{\alpha_{n-1}\over 2!} + \cdots + {\alpha_0\over (n+1)!} = 0,
\quad\quad\quad {\rm for}\ \ n=1,2,\cdots,   \label{alpha}
\end{eqnarray}
as
\begin{eqnarray}
\alpha_0 =1,\quad \alpha_1=-{1\over 2},\quad \alpha_2 ={1\over 12}, \quad
\alpha_3=0, \quad \alpha_4=-{1\over 720}, \quad\cdots\cdots. \nonumber
\end{eqnarray}
(\ref{formula1})$\sim$(\ref{alpha})  will be proved at the end of this appendix. 
In (\ref{non-def1}) $(ad\ i\phi\cdot T)^n$ is a mapping defined by the  $n$-ple commutator
\begin{eqnarray}
(ad\ i\phi\cdot T)^n O=[i\phi\cdot T,\cdots,[i\phi\cdot T,[i\phi\cdot T,O]]\cdots].   \label{multi}
\end{eqnarray}
As has been discussed in subsection 5.1, the grading of the commutator may differ  depending on the 
representation. We employed the prescription  (\ref{Pres1}) for the unitary(oscillator) representation,
 but the one (\ref{Pres2}) for the matrix representation. For the respective case (\ref{multi}) reads
$$
i^n M^\Xi\phi^{\bar i_{n}}\phi^{\bar i_{n-1}}\cdots \phi^{\bar i_{1}}
[T^{i_{1}},\cdots,[T^{i_{n}},T^{\Xi}\}\}\cdots\}
$$
or
$$
i^n\phi^{\bar i_{1}}\phi^{\bar i_{2}}\cdots \phi^{\bar i_{n}}M^\Xi
[T^{i_{1}},\cdots,[T^{i_{n}},T^{\Xi}\}\}\cdots\}.
$$
Here $[\ , \}$ can be automatically read as a commutator or anti-commutator from  the grading  of $\phi^{\bar i_{1}},\phi^{\bar i_{2}},\cdots$. Note that the difference of the two prescriptions  reflects merely on the ordering  of $\phi^{\bar i}$s. For a simple case where $n=1$ and $M^I=0$ it reads 
\begin{eqnarray}
(ad\ \phi\cdot T) (T^\Xi M^\Xi){\Big |}_{M^{I}=0}&=& X^{{\overline M}_1} M^{{\overline M}_2}[T^{{\overline M}_1},T^{{\overline M}_2}]+X^{{\overline M}_1}M^{{\overline {\frak M}}_2}[T^{{\overline M}_1},T^{\overline{\frak M}_2}]   \nonumber\\
&+&\Theta^{{\overline {\frak M}}_1}M^{{\overline  M}_2}[T^{{\overline {\frak M}}_1},T^{{\bar M}_2}] \pm\Theta^{{\overline {\frak M}}_1}M^{{\overline {\frak M}}_2}\{T^{{\overline {\frak M}}_1},T^{{\overline {\frak M}}_2}\},   \nonumber
\end{eqnarray}
with the sign taken to be $+$ for the unitary(oscillator) representation, but - for the matrix representation. 
 Here use was made of the index notation (\ref{rename}) for $\phi^{\bar i}$.

We expand  $R^{\Xi}(\phi)$ and  $\rho^\Xi(\phi)$  in series of  $\phi^{\bar i}$:
\begin{eqnarray}
R^{\Xi}(\phi) &=& R^{\Xi}_{(0)}(\phi) + R^{\Xi}_{(1)}(\phi)+ \cdots + R^{\Xi}_{(n)}(\phi)+ 
\cdots, 
\nonumber\\
\rho^\Xi(\phi,M) &=&\rho^\Xi_{(0)}(\phi,M) +\rho^\Xi_{(1)}(\phi,M)+ \cdots+ \rho^\Xi_{(n)}(\phi,M)+ \cdots.   \nonumber
\end{eqnarray}
 The $n$-th order terms $R^{\Xi}_{(n)}(\phi)$ and $\rho^\Xi_{(n)}(\phi,M)$ of the expansion  obey the  recursive relations obtained by comparing the powers of both sides of (\ref{non-def1}) order by order. We find 

\vspace{0.5cm}
\noindent 
to the $0$-th order of $\phi$
\begin{eqnarray}
R^\Xi_{(0)}(z)\cdot T + \alpha_0\biggl(\rho^\Xi_{(0)}(\phi,M)\cdot H \biggr) = i\alpha_0T^\Xi,   \nonumber
\end{eqnarray}

\noindent 
to the first order of  $\phi$
\begin{eqnarray}
R^\Xi_{(1)}(\phi)\cdot T +\alpha_0\biggl(\rho^\Xi_{(1)}(\phi,M)\cdot H \biggr) =i\alpha_1\biggl[\phi\cdot T,\ T^\Xi\biggr] 
- \alpha_1\biggl[\phi\cdot T,\ \rho^\Xi_{(0)}(\phi,M)\cdot H \biggr], \ \ \nonumber
\end{eqnarray}

\noindent
to the second order in $z$
\begin{eqnarray}
R^\Xi_{(2)}(\phi)\cdot T &+&\alpha_0\biggl(\rho^\Xi_{(2)}(\phi,M)\cdot H \biggr) 
= i\alpha_2\biggl[\phi\cdot T,\biggl[\phi\cdot T,\ T^\Xi\biggr]\biggr] 
- \alpha_1\biggl[\phi\cdot T,\ \rho^\Xi_{(1)}(\phi,M)\cdot H \biggr], \nonumber
\end{eqnarray}
and so on. 
 These recursion relations are solved for $R^{\Xi \bar i}_{(n)}(\phi)$ and $\rho^\Xi_{(n)}(\phi,M)$ order by order, by setting  $R^{\Xi}_{(0)}(\phi)=i\alpha_0\delta^{\Xi \bar i}$ as an initial condition.  Thus the Killing vectors are determined once the algebra of $\{T^\Xi\}$ given with the decomposition (\ref{decompo}). 

 Finally we will prove (\ref{formula1}). (\ref{formula2}) can be proved similarly. 
First of all, from  the Hausdorff formula 
\begin{eqnarray}
e^Xe^Y= \exp\Big[\sum_m\sum_{p_i\ge 0,p_j\ge 0 \atop p_i+q_i>0} {(-1)^n\over m}{(ad X)^{p_1}(ad Y)^{q_1}\cdots\cdots(ad X)^{p_n}(ad Y)^{q_{n-1}}Y \over p_1!q_1!\cdots\cdots p_n!q_n!
(p_1+q_1+\cdots\cdots+p_n+q_n)}\Big].  \nonumber
\end{eqnarray}
 we know that (\ref{formula1}) holds with some coefficients $\alpha_n, n=0,1,2,\cdots$ if ${\cal E} \ll 1$. To determine the coefficients we expand the exponents of both sides. The expansion of the l.h.s. is simply 
\begin{eqnarray}
l.h.s =(1+{\cal E})\sum_{n=0}^\infty {1\over n!}X^n.  \nonumber 
\end{eqnarray}
But we expand   the r.h.s.  in $X$ and ${\cal E}$,  retaining the terms of $O({\cal E})$:
\begin{eqnarray}
r.h.s &=&  \sum_{n=0}^\infty {1\over n!}
\biggl( X+\sum_{n=0}^\infty \alpha_n(ad\ X)^n{\cal E}  \biggr)^n + 
O({\cal E}^2)  \nonumber \\
&=& 1 + \biggl( X+\sum_{n=0}^\infty \alpha_n(ad\ X)^n{\cal E}  \biggr) 
 \nonumber \\
&\quad& + {1\over 2!}\biggl( X^2 + X\sum_{n=0}^\infty \alpha_n(ad\ X)^n{\cal E}  + \sum_{n=0}^\infty \alpha_n(ad\ X)^n{\cal E}\cdot X\biggr)  
\nonumber \\
 + {1\over 3!}\biggl( X^3 &+& X^2\sum_{n=0}^\infty \alpha_n(ad\ X)^n{\cal E}  + X\sum_{n=0}^\infty \alpha_n(ad\ X)^n{\cal E}\cdot X  
 + \sum_{n=0}^\infty \alpha_n(ad\ X)^n{\cal E}\cdot X^2 \biggr)  \nonumber\\
 &\ & \hspace{3cm} + O({\cal E}^2)  \nonumber
\end{eqnarray}
Expanding this  in $X$ furthermore  we find  

\vspace{0.5cm}
\noindent 
to the  $0$-th oder in $X$:   
$$
  1+\alpha_0{\cal E}, 
$$

\noindent 
to the first order in $X$:
\begin{eqnarray}
 (1+\alpha_0{\cal E})X + (\alpha_1 + {\alpha_0\over 2!})(ad\ X),\nonumber  
\label{order1}
\end{eqnarray}

\noindent 
to the second order in $X$:
\begin{eqnarray}
 {1\over 2!}(1+&\alpha_0{\cal E}&)X^2 +
(\alpha_2+{\alpha_1\over 2!}+{\alpha_0\over 3!})(ad\ X)^2{\cal E}
+ (\alpha_1+{\alpha_0\over 2!})(ad\ X){\cal E}\cdot X, \nonumber
\end{eqnarray}

\noindent 
to the third order in $X$:  
\begin{eqnarray}
 {1\over 3!}(1+&\alpha_0{\cal E}&)X^3 +
(\alpha_3+{\alpha_2\over 2!}+{\alpha_1\over 3!}+{\alpha_0\over 4!})(ad\ X)^3{\cal E} \nonumber \\
&+& (\alpha_2+{\alpha_1\over 2!}+ {\alpha_0\over 3})(ad\ X)^2{\cal E}\cdot X
+ {1\over 2!}(\alpha_1 + {\alpha_0\over 2!})(ad\ X){\cal E}\cdot X^2. \nonumber
\label{order3}
\end{eqnarray}
If $\alpha_0=1$ and the recursion relation (\ref{alpha}) holds up to $n=3$, then both sides of (\ref{formula1} ) are equal up to the third order of $X$. 
 The calculation  may  be inductively generalized  to any  order of $X$. 
Thus (\ref{formula1} ) was proved.

\end{document}